\documentclass[twocolumn]{aastex701}
\usepackage{float}
\usepackage{xcolor}

\usepackage{amsmath}
\usepackage{gensymb}
\usepackage{graphicx}
\usepackage{hyperref}
\newcommand{\R}{\ensuremath{R(V)}}
\newcommand{\E}{\ensuremath{E(B-V)}}

\begin{document}

\title{Optical-Infrared Extinction Variation in Nearby Molecular Clouds}

\author[orcid=0009-0000-4365-0204]{Marshall Hobson-Ritz}
\affiliation{University of Maryland, College Park, MD, USA}
\affiliation{Space Telescope Science Institute, Baltimore, MD, USA}
\email{mhobritz@umd.edu}

\author[orcid=0009-0005-7982-2369]{Sarv Parteek Singh}
\affiliation{John Hopkins University, Baltimore, MD, USA}
\email{ssingh60@jh.edu}

\author[orcid=0000-0002-3569-7421]{Eddie Schlafly}
\affiliation{Space Telescope Science Institute, Baltimore, MD, USA}
\email{eschlafly@stsci.edu}

\author[orcid=0009-0005-7982-2369]{Lee Mundy}
\affiliation{University of Maryland, College Park, MD, USA}
\email{lgm@umd.edu}

\author[orcid=0000-0002-5480-5686]{Alberto Bolatto}
\affiliation{University of Maryland, College Park, MD, USA}
\email{bolatto@umd.edu}

\begin{abstract}
The dust extinction curve is an important probe of the properties of dust grains, particularly their size distribution. The properties of dust grains are expected to change significantly in dense molecular clouds due to shielding from interstellar radiation and increased frequency of grain-grain and grain-gas collisions. We measure the dust extinction curve shape parameter \R\ in 9 nearby, dense molecular clouds using data from the APOGEE spectroscopic survey, in combination with ten-band photometry from PS1, 2MASS, and WISE. We analyze the results to determine how the extinction curve varies in nearby molecular clouds. We find that most molecular clouds show a positive correlation between \R\ and \E, but the strength of the correlation varies significantly between clouds. In addition, we find that the clouds with the strongest correlation tend to have very high dispersion in \R\ at fixed \E, which tightens dramatically when clouds are subdivided into separate regions. This suggests that local environmental conditions beyond density shape the $\R-\E$ relation.

\end{abstract}


\section{Introduction} \label{sec:intro}

Dust is a basic ingredient of star formation, and is composed of small, solid grains of material. Individually, grains can vary in shape, size, and composition. Although these properties are initially set at formation, they can change through several processes including accretion of gas-phase material of the interstellar medium (ISM) \citep{Draine2009}, formation of ices on the mantles \citep{Hulst_1946}, and coagulation \citep{Ormel_2009}. On large grains, more complex surface chemistry can occur \citep{TH_1982}, with the caveat that these ices can only form at low temperatures. Molecular clouds are the only regions cold enough to enable ice formation, often associating larger grains with sites of star formation within molecular clouds. One of the best ways to probe grain sizes, is to look at the extinction of background stars.

Due to their small size, dust grains preferentially scatter and absorb blue light relative to red light. This relationship between extinction and wavelength is called the dust extinction curve \citep{Draine_2003}, and was parameterized by \cite{FM_1986,FM_1988} in the optical and ultraviolet (UV). \cite{Cardelli_1989} found much of the variation could be described by the parameter \R, which describes the total-to-selective extinction ratio. Here $\R=A(V)/\E$, where $A(V)$ is the extinction in the $V$ band and \E\ is the excess extinction of the $B$ to $V$ band, which is proportional to the dust column density \citep{Savage_1977,Bohlin_1978}. These diagnostics for the shape of the extinction curve help provide insight into dust grain properties. Through these diagnostics, it is known that there is widespread variation of \R\ across the Galaxy \citep{Fitzpatrick_2004,Schlafly_2016,Zhang_2025,beyond_rv_2025}.

This variation in \R\ is often attributed to the size distribution of grains. Dust grains effectively extinguish light with wavelengths smaller than the size of the grain, so as the average grain size increases, the ratio of light extinguished at large wavelengths relative to small wavelengths increases. This corresponds to a flatter extinction curve and higher \R, giving rise to the association of \R\ with grain size. The classic interpretation is that grains are able to grow in denser regions ($\E>1.0$), where they are shielded from external radiation. An alternative idea is that the variation may instead be due to ultraviolet photons processing the smaller grains or the composition and chemistry of the grains themselves \citep{Jones_2013,Mulas_2013}. \cite{Bohlin_1978} also noted that higher-density sight lines lie above lower-density ones in the N(H) vs \E\ relation, likely reflecting changes in the extinction curve due to the growth of small grains into larger ones. These early observations suggest that dense regions may exhibit systematic variations in extinction that are not captured by broad Galactic surveys.

The work of \cite{Schlafly_2016} (hereafter S16) measured \R\ for tens of thousands of stars with spectroscopy from APOGEE \citep{APOGEE_1_2017} and photometry from PS1, 2MASS, and WISE \citep{PS1_database,2MASS,allwise}. The limited variations in \R\ in their sample ($\sigma_{\R} = 0.18$) showed strong correlations over large spatial scales (i.e. the Galactic radius) \citep{Schlafly_2017}. Additionally, they found no correlation between \R\ and \E\ in their sample, which largely focused on $\E < 2$. That work, however, focused on the broad statistical properties of the extinction curve and did not investigate any particular molecular clouds in detail.

A study similar in scope, by \cite{Zhang_2025}, measured \R\ for 130 million stars with low-resolution Gaia XP spectra \citep{Gaia_DR3_proc_spec, Gaia_DR3_spec_calib}. This work similarly found variations in \R\ on the scale of the Galactic radius, and also examined variations within the regions of molecular clouds. While this work found variations in the $\R-\E$ relationship in these clouds, the Gaia data are limited at the highest extinctions ($\E \sim2.0$), restricting sensitivity to regions of lower extinction.

Our work takes a similar approach to S16, but uses more recent observations and focuses closely on nearby molecular clouds. They are precisely the regions we expect to see the most dramatic variations in the extinction curve, due to the high densities and varied radiation fields. Moreover, they are the site of star formation, with the nearest clouds being among the best studied regions in astronomy.

In this paper, we examine \R\ trends in nearby molecular clouds, with the aim of characterizing how the extinction curve varies, and connecting that variation to the growth of grains within individual clouds. We additionally investigate to what extent these trends are universal. Our work extends previous analyses to higher extinctions, $\E \gtrsim2.0$, allowing us to probe more deeply embedded regions. Alongside this analysis, we provide a catalog of \R\ and \E\ proxy values for over 50,000 stars in the Milky Way for future use in extinction studies.

This paper is organized as follows. In \textsection\ref{sec:data}, we discuss the observations and outline our data selection process. In \textsection\ref{sec:methods} we describe the details of the technique applied, how we compare to other \R\ measurements, and our methodology for selecting our molecular cloud sample. In \textsection\ref{sec:MC_ext} we provide an overview of the catalog, and walk through our analysis of our cloud sample on a high level. In \textsection\ref{sec:drivers} we discuss these results to try to understand mechanisms driving extinction, examining patterns collectively and on an individual cloud basis. Finally, in \textsection\ref{sec:conclusion} we summarize and conclude.

\begin{figure*}
    \centering
    \includegraphics[width=\linewidth]{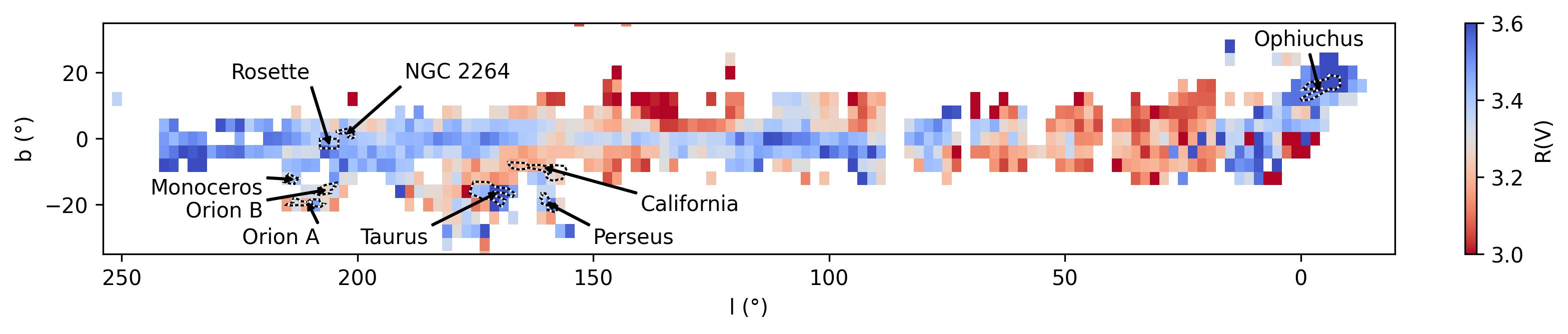}
    \caption{A map of \R, related to the slope of the extinction curve, as measured in this work, using 52,912 stars. We focus in this work particularly on the stars in nearby molecular clouds (labeled), where extinction variations tend to be more dramatic than more generally throughout the Milky Way. \textsection \ref{sec:param_ext_curve} describes how we find \R.}
    \label{fig:sky_coverage}
\end{figure*}

\section{Data} \label{sec:data}

This work uses observations from a variety of surveys to measure the shape of the extinction curve and to correlate those measurements with other probes of the interstellar medium.

    \subsection{Spectroscopy}
    Our observational footprint is defined by the Apache Point Observatory (APO) Galactic Evolution Experiment (APOGEE). APOGEE is a spectroscopic survey utilizing two 2.5m telescopes, one at APO and one at Las Campanas Observatory (LCO). The survey maps the chemical and dynamical patterns of stars in the Milky Way through high resolution ($R>22,500$) spectroscopy in the near IR. The spectrographs on both telescopes utilize 300 fibers to take spectra in the H-band ($1.51 \mu m < \lambda < 1.70 \mu m$), with a high signal-to-noise ratio ($100> pixel^{-1}$) for most targets \citep{apogee_2015}. The second telescope at LCO was added for the APOGEE-2 survey, which continues the work from APOGEE-1 with additional targeting in the southern hemisphere. APOGEE primarily targets red clump and red giant stars, which are selected based on the de-reddened $J-K_S$ color, and can observe down to a limiting magnitude of 12.2-13.8. We note that this work uses primarily spectra from APO, due to our reliance on PS1 photometry which is only available for $\delta > -30\degree$.
    
    The ASPCAP pipeline determines the temperature, metallicity, surface gravity, and detailed chemical abundances of APOGEE stars \citep{apo_param_pipe,apo_spec_pipe}. We use APOGEE data from Sloan Digital Sky Survey (SDSS) IV data release 17 (DR17) \citep{SDSS4}, which has spectra for over 400,000 stars. We note that ASPCAP operates on continuum-normalized infrared spectra, so that the derived stellar temperatures, metallicities, and gravities are not sensitive to the extinction of the stars.

    This work also includes ancillary data from the APOGEE Reddening Survey, which targeted Milky Way stars behind molecular clouds in the solar neighborhood (Ophiuchus, Orion, Perseus, and Taurus). This ancillary survey followed the general target selection algorithm of the main APOGEE sample \citep{apogee_targeting}, with the additional requirement the estimated photometric reddening $A(K)$ to each target be greater than 0.3~mag for stars behind clouds in the solar neighborhood. This work includes over 300 targets from the APOGEE Reddening Survey.
    
    We use the measurements of the effective temperature, metallicity, and gravity of APOGEE stars to calculate their intrinsic colors, setting the footprint for our dataset coverage. Additionally, we use parallax and parallax error from Gaia data release 3, which are provided in DR17 \citep{gaia, gaia_summary}.
    
    \subsection{Photometry}
     To measure the extinction curve, we use several surveys to measure the apparent magnitude of the stars in APOGEE for different parts of the electromagnetic spectrum.
     
     For optical photometry, we use the Panoramic Survey Telescope and Rapid Response System (PS1). Observations are made with the $g_{P1}$ (480~nm), $r_{P1}$ (620~nm), $i_{P1}$ (750~nm), $z_{P1}$ (870~nm), and $y_{P1}$ (960~nm) filters, with limiting magnitudes of 23.2, 23.2, 23.1, 22.3, and 21.2. The final products are available in the PS1 database \citep{PS1_database}. For infrared (IR) photometry, we use the Two Micron All Sky Survey (2MASS) \citep{2MASS}. The survey measured photometry in the J (1.25~$\mu m$), H (1.65~$\mu m$), and $K_S$ (2.16~$\mu m$) bands, with respective limiting magnitudes of 15, 14.3, and 13.5. For mid-IR photometry, we use the Wide-field Infrared Survey Explorer \citep[WISE;][]{wise}. The survey measured photometry in the W1 (3.4~$\mu m$), W2 (4.6~$\mu m$), W3 (12~$\mu m$), and W4 (22~$\mu m$) bands, with respective limiting magnitudes of 16.5, 15.5, 11.2, and 7. This work uses photometry from ALLWISE \citep{allwise}, containing data from two different sky coverage epochs.
     
     We limit our use of ALLWISE to the W1 and W2 filters, which maintained full sensitivity across the epochs. WISE observed over 700 million objects, with the stipulation that several have been flagged as unreliable photometry due to an overlap with diffraction spikes or a bright nearby star. For these flagged WISE sources, we assign an infinite photometric uncertainty.
    
    \subsection{Target Selection}
    We adopt selection criteria similar to those used by S16. Specifically, we (1) limit the sample to targets with PS1 $y_{P1}<13$ mag, (2) exclude APOGEE ``ancillary'' targets other than targets from the APOGEE Reddening Survey, (3) remove stars without reliable surface gravity measurements or with APOGEE flags indicating abnormally high $\chi^2$ values or rapid rotation, and (4) exclude any objects whose PS1 or WISE positions differ from their 2MASS positions by more than 0.5".

    We then apply the following additional selection criteria:
    \begin{enumerate}
        \item Limit targets to those with $\E>0.3$. Below this threshold, the uncertainty in \R\ is roughly comparable to the typical variation in \R\ \citep{Schlafly_2016}.

        \item Remove targets classified as young stellar objects (YSOs) from their placement on the Kiel diagram. We use the following empirical relation to remove YSOs from the sample: $$log\ g < 1.3 + (2.6/1500)(T_{eff}-3500).$$ Where g is the surface gravity and $T_{eff}$ is the effective temperature  of the stars.

        \item Require targets to have photometric uncertainties less than 0.2 mag in at least seven bands. This ensures that we use only stars for which good measurements of \R\ are possible.

        \item Exclude targets whose parallax measurements indicate they are within, or in front of, the molecular cloud. We want all of our stars to be clearly behind the cloud so that we can compare the extinction measurements with other tracers of the total column.
    \end{enumerate}
    
    Figure \ref{fig:sky_coverage} displays our derived \R\ for the final selected stars over the sky. The stars cover most of the Galactic plane with $\delta > 30\degree$, including our targeted molecular clouds.

\begin{table*}
    \centering
    \begin{tabular}{lccccccccc}
    \hline
    \hline
         Cloud & $l$ range [$\degree$]& $b$ range [$\degree$] & $D$ [pc] &  $N_{stars}$ & $\langle\E\rangle$& $\langle\R\rangle$ & $\sigma_{\E}$ & $\sigma_{R(V)}$ & Corr[\R, \E]\\
    \hline
         Taurus & 166, 178 & -21, -9 & 155 &  299 & 1.13 & 3.16 & 0.65 & 0.33 & Strong\\
         
         Perseus & 155, 165 & -25, -15 & 325 & 153 & 0.99 & 3.09 & 0.45 & 0.20 & Strong\\
         
         Orion A & 207, 216 & -25, -16 & 475 & 34 & 1.52 & 3.00 & 0.88 & 0.14 & Weak\\
         
         Orion B & 203, 210 & -23, -13 & 475 & 120 & 1.46 & 2.99 & 0.53 & 0.18 & Weak\\
         
         Monoceros & 212, 216 & -14, -10 & 860 & 99 & 0.93 & 3.02 & 0.37 & 0.15 & Moderate\\

         NGC 2264 & 200.5, 204.5 & 0, 4 & 820 & 56 & 1.01 & 3.15 & 0.44 & 0.14 & Moderate\\

         Ophiuchus & 350, 360 & 10, 20 & 160 & 455 & 0.72 & 3.44 & 0.25 & 0.27 & Strong\\

         California & 155, 168 & -18, -5 & 520 & 379 & 0.80 & 2.92 & 0.22 & 0.11 & Weak\\

         NGC 1579 & 164.5, 166 & -9.5, -8.0 & 520 & 79 & 0.82 & 2.91 & 0.25 & 0.10 & Strong\\

         Rosette & 204, 208 & -4, 0 & 1440 & 506 & 0.87 & 3.17 & 0.30 & 0.15 & Moderate\\
    \hline
    \end{tabular}
    \caption{The ranges of galactic longitude (l) and latitude (b), distance to the far edge of the cloud ($\sim1.1\times$\cite{cloud_dist_zuker} mean distances), number of stars ($N_{stars}$), median \E, median \R, median absolute deviation of \E, median absolute deviation of \R\ for a given cloud, and our categorization of of the strength of correlation between \R\ and \E. Where \E\ and \R\, in these clouds, have typical errors of 0.004 mag and 0.079 respectively. The ranges of l and b provide the frame the cloud is presented in, but the remaining column values are from the stars that are within the spatial mask, described in section \ref{sec:only_cloud}.}
    \label{tab:clouds}
\end{table*}
\section{Methods} \label{sec:methods}

    \subsection{Parameterizing Extinction Curves}\label{sec:param_ext_curve}
    To find the extinction properties of our sample, we largely follow the methods outlined in S16. We model each star's intrinsic magnitudes using the empirical relationship from S16 to fit the metallicity and temperature across ten observational bands. The comparison between our observed and intrinsic magnitudes allows us to infer the line-of-sight extinction parameters by solving the following system:
        $$\textbf{m}_{obs} - \textbf{m}_{int} = \textbf{Ay},$$    
    where $\textbf{m}_{obs}$ and $\textbf{m}_{int}$ are the vectors of the 10 bands of observed and intrinsic magnitudes respectively.
    
    The matrix \textbf{A} describes the effect of changing the distance, reddening, and extinction curve on stars, and $\textbf{y}$ gives the particular coefficients of the parameters in \textbf{A} corresponding to individual stars. In $\textbf{A}$, each row of the 10 by 3 array corresponds to a photometric band, with the columns:
        $$\textbf{A} = \left(
        \begin{array}{ccc}
             \uparrow & \uparrow & \uparrow\\
             \bf1 & \bf R_0 & \bf dR/dx\\
             \downarrow & \downarrow & \downarrow\\
        \end{array}\right),$$
     where we use the $R_0$ and $dR/dx$ values provided in S16. Separately, $\textbf{y} = [\mu, E', E'x]$; where, $\mu$ is the distance modulus, E' represents the dust column, and x describes the shape of the extinction curve.
    
    We convert E' and x into proxies for \E\ and \R\ with the following transformations from S16:
        \begin{equation}
            \E = E'\times(R_{0,g}-R_{0,r})\times 1.02,
            \label{eq_E}
        \end{equation}
        \begin{equation}
            \text{ and } \R\ = 9.1x+3.3.
            \label{eq_RV}
        \end{equation}
    E' is not normalized to any bands, this multiplication with $(R_{0,g}-R_{0,r})$ re-normalizes to the g and r bands. The factor of 1.02 corresponds to the $E(B-V)/E(g-r)$ value for the F99 reddening law \citep{fitzpatrick99}. It is worth noting this ratio is very close to 1.0, indicating that extinction measurements made in the g and r bands are very similar to those made in the the B and V bands. With this and the similarity of A(V) to $E(g-W2)$, S16 found the linear relationship of $E(g-W2)/E(g-r)$ to \cite{fitzpatrick99} \R\ predictions, resulting in Equation \ref{eq_RV}. Although this method is not an explicit measurement of \R\ or \E, the conversion follows the \cite{fitzpatrick99} predictions very closely. We compute $\mu$, \E, \R, and the associated uncertainties using standard linear algebra and propagation of uncertainties (Appendix \ref{app:methods}).

    \subsection{Target Selection Within Molecular Clouds}\label{sec:only_cloud}
    We choose to limit our analysis to stars in the main bodies of nearby molecular cloud. In this work we utilize extinction maps in the J-band, derived from 2MASS observations and the Near-Infrared Color Excess Revised (NICER) and NICEST technique \citep{Juvela_NICERST}. We manually defined polygons around each of the molecular clouds in our sample using \texttt{SAO DS9} and the $A(J)$ maps for each cloud (examples can be seen in the first 3 panels of Figures \ref{fig:Taurus4}-\ref{fig:California4}). In the cases of the Rosette Nebula and NGC 1597, we simply included all of the stars in a large rectangular region around the cloud because our selection contained no stars outside the main body of the clouds. These selections allow us to focus on regions of high extinction.
            
    \begin{figure}
        \centering
        \includegraphics[width=\linewidth]{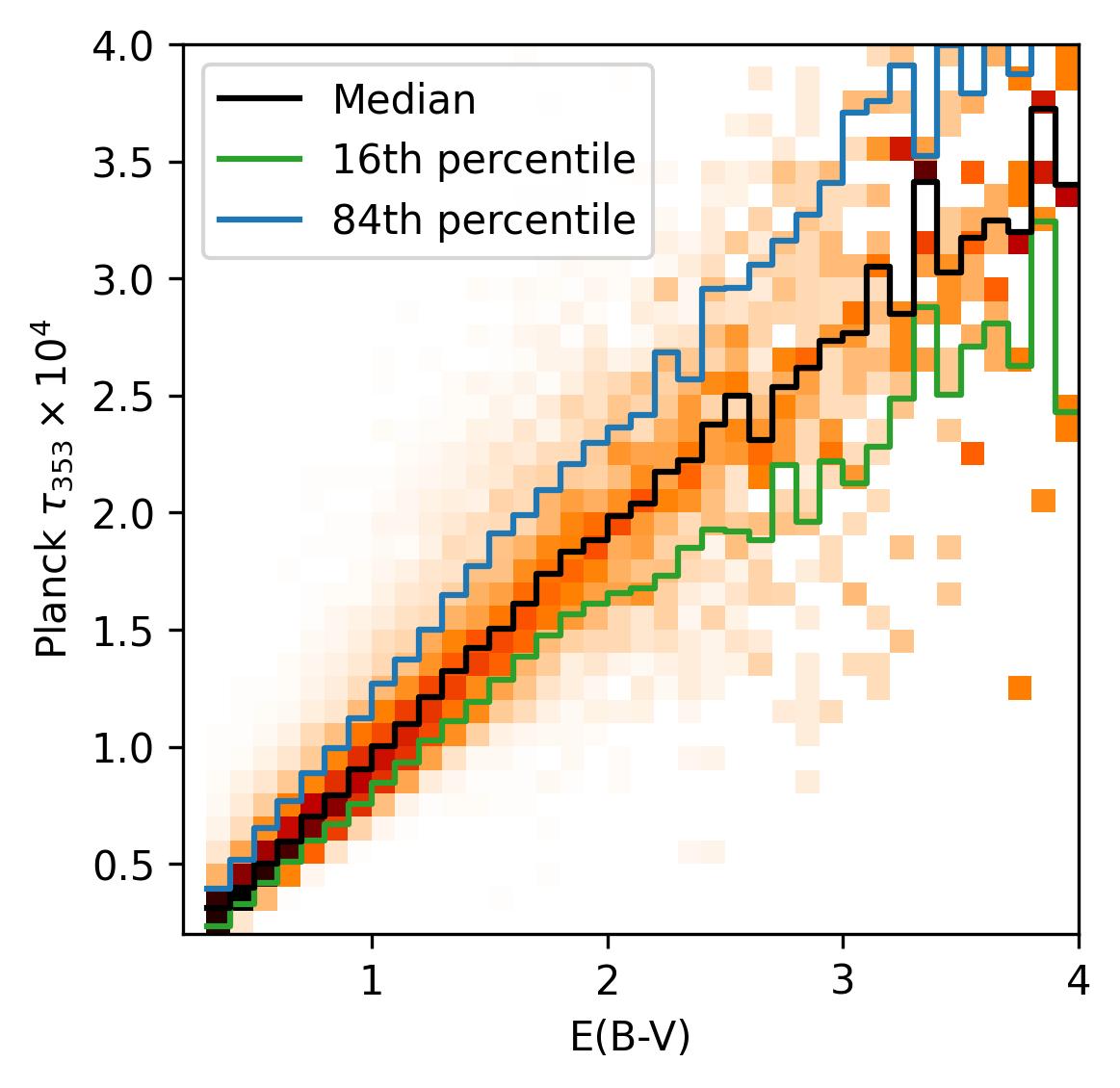}
        \caption{Comparison between this work's \E\ measurements and the \textit{Planck} optical depth at 353GHz ($\tau_{353}$), scaled on the y-axis by a factor of $10^4$. The black line is the binned median and the green and blue lines are the 16th and 84th percentile respectively.}
        \label{fig:planck_tau_comp}
    \end{figure}
 
    \begin{figure}
        \centering
        \includegraphics[width=\linewidth]{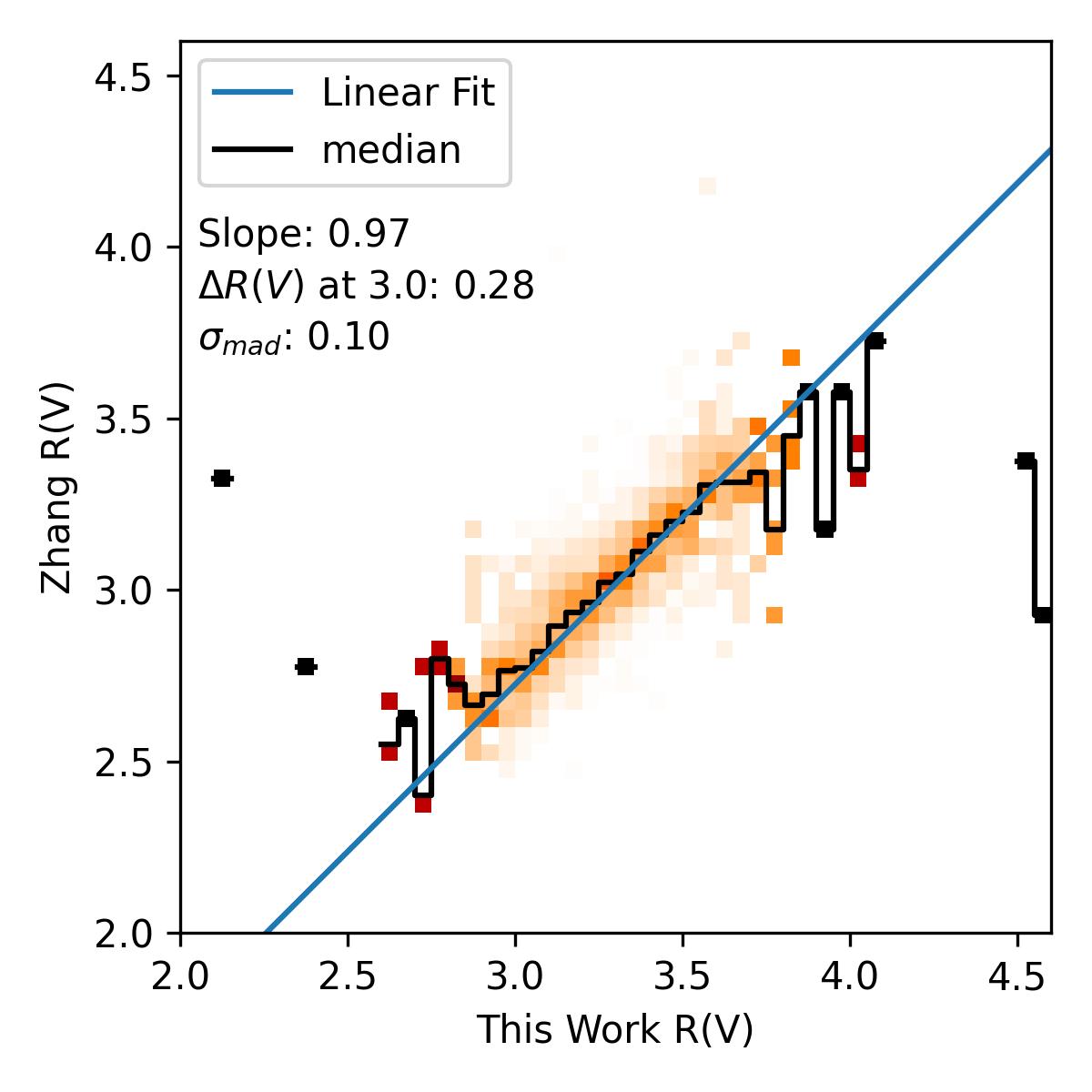}
        \caption{A 2D histogram of \R\ measurements of this work on the x-axis and from Z25 on the y-axis. The black line shows the median value in each bin, and the blue line shows our linear fit between the two measurement systems. The upper left corner of the plot displays information about this fit.}
        \label{fig:ZvS_PS}
    \end{figure}

\section{Extinction in Nearby Molecular Clouds} \label{sec:MC_ext}
    \subsection{Overview of Catalog}
    Our catalog includes measurements of \E, and \R, and their uncertainties, for 52,912 stars. We restrict our analysis to molecular clouds located more than $10^\circ$ from the Galactic center, minimizing contamination from additional background Galactic extinction. We limit our work to clouds containing stars where $\E>1$, to focus on clouds with dense, highly extinct regions. Table \ref{tab:clouds} lists the clouds in our sample along with the coordinates, distances, and other relevant properties for our analysis. Our catalog is publicly available on \href{https://zenodo.org/records/22117631}{zenodo}.

    \subsection{Verification of Measurements}
    To verify that our extinction measurements behave as expected, we compare the \E\ values to the \textit{Planck} \citep{Planck_2014} optical depth at 353GHz ($\tau_{353}$). We first filter our measurements to be those with 10 bands of measurements with photometric errors$<$0.2 in all 10 bands, and to be outside the thickest region of the galactic disk ($l<60$ and $|b|<5$). For each measurement, we interpolate $\tau_{353}$ value on the stellar coordinates. Figure \ref{fig:planck_tau_comp} shows our \E\ extinction measurements have an approximately linear relationship with $\tau_{353}$, where the percentiles widen as the number of measurements decreases at higher dust column. This linear increase with extinction measurements indicates \E\ is working as anticipated.

    With establishing that \E\ measurements behave as expected, the next behavior to check is that of \R. \cite{Zhang_2025} (hereafter Z25) measured \R\ for hundreds of millions of stars by combining low-resolution Gaia XP spectroscopy with sophisticated data-driven Bayesian models of stellar spectra and the dust extinction curve. This \R\ map has full sky coverage, making an ideal point of comparison. We directly compare our \R\ values with those from Z25 by matching the results for stars included in both catalogs, we use a match threshold of one arcsecond.
    
    With matching stars, we compute a linear fit between the two sets of \R\ measurements. Figure \ref{fig:ZvS_PS} shows the result, with a slope of 0.97 and $\sigma_{mad}=0.10$, indicating very good agreement between the measurements. However, the two sets of measurements have an offset of $\Delta$\R\ of 0.27 at $R(V) = 3.0$. An overall constant offset between different \R\ measurements is in some way expected; for example, S16 shows that the choosing between the F99 and CCM extinction curves can lead to different predictions of \R\ by up to $\sim0.5$ that is largely due to an offset in models with similar slopes. In this work we focus on the variation in the extinction curve, so we therefore choose to apply this linear fit: $$\R_\mathrm{Zhang} = 0.974 \cdot \R_\mathrm{this\ work}-0.197,$$ to put our values on the Z25 \R\ scale, inheriting their average \R.

\begin{figure}
        \centering
        \includegraphics[width=\linewidth]{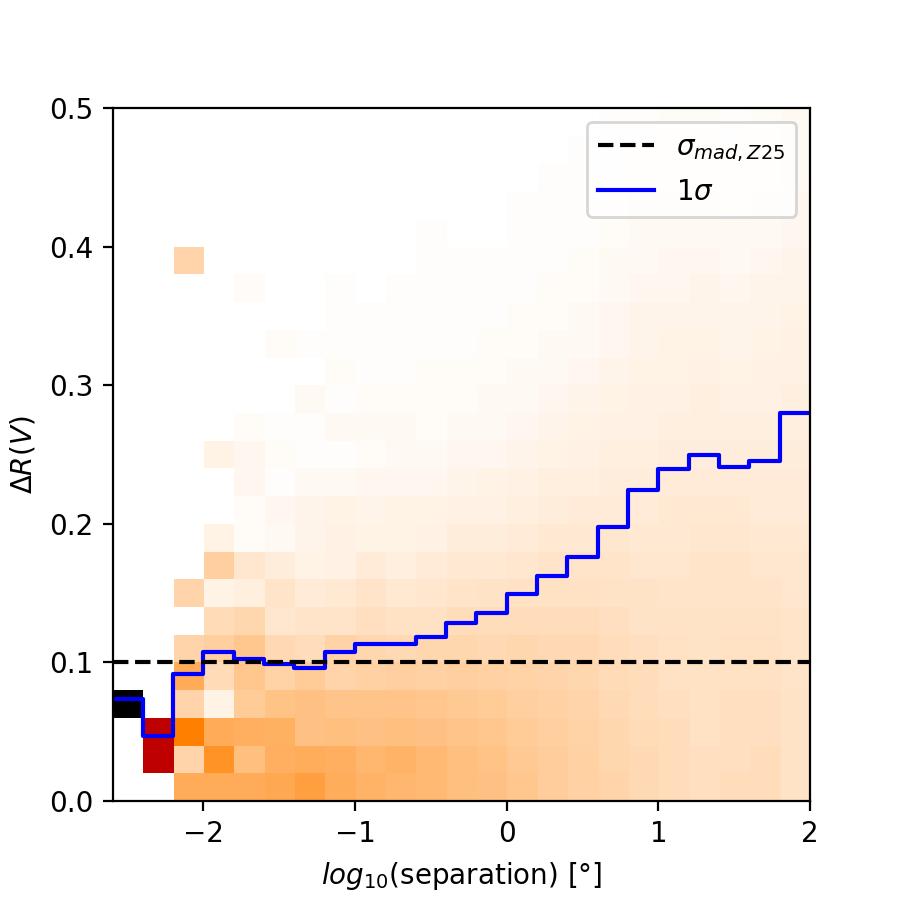}
        \caption{A 2D histogram of the difference of \R\ versus the separation of two stars. The dashed black line shows the $\sigma_{mad}$ from the comparison with Z25 in Figure \ref{fig:ZvS_PS}. The blue line displays, at each separation bin, the 1$\sigma_{\R}$ value.}
        \label{fig:nearest_neighbors}
    \end{figure}

    With \E\ and \R\ behaving as expected, we now check our 1$\sigma_{\R}$ errors through nearest neighbor analysis. We first filter our measurements to be those with 10 bands of measurements with photometric errors$<$0.2 in all 10 bands, to be outside the thickest region of the galactic disk ($l<60$ and $|b|<5$), and have measurements within ranges of \E\ and \R\ that are expected to be well characterized ($0.8<\E<3.0$ and $2<\R<5$). Figure \ref{fig:nearest_neighbors} shows how the difference in \R\ between two stars changes as the separation increases. For comparison, the $\sigma_{mad}$ from our comparison to Z25 is shown as a horizontal black line. We find that the approximate $1\sigma_{\R}$ at small 2D distances is $\sim0.1$, consistent with $\sigma_{mad,Z25}$. As stars become closer, it is expected that $(\Delta \R)^2 \sim 2\sigma_{\R}$, with a resulting $\langle\sigma_{\R}\rangle\sim0.07$. The by-star 1$\sigma_{\R}$ values are, on average, 0.05. It is unclear what additional source of uncertainty causes this discrepancy. One possibility is that our method currently only considers photometric errors, where there are additional spectroscopic parameter errors we do not account for. Understanding our error measurements are likely smaller than the true errors, we add in quadrature $\frac{0.1}{\sqrt{2}}$ for the reported $\sigma_{\R}$, creating a floor value of 0.07. Regardless of this correction, the $\sigma_{\R}$ values are much smaller than the \R\ variations we investigate in this work.

    \begin{figure*}
        \centering
        \includegraphics{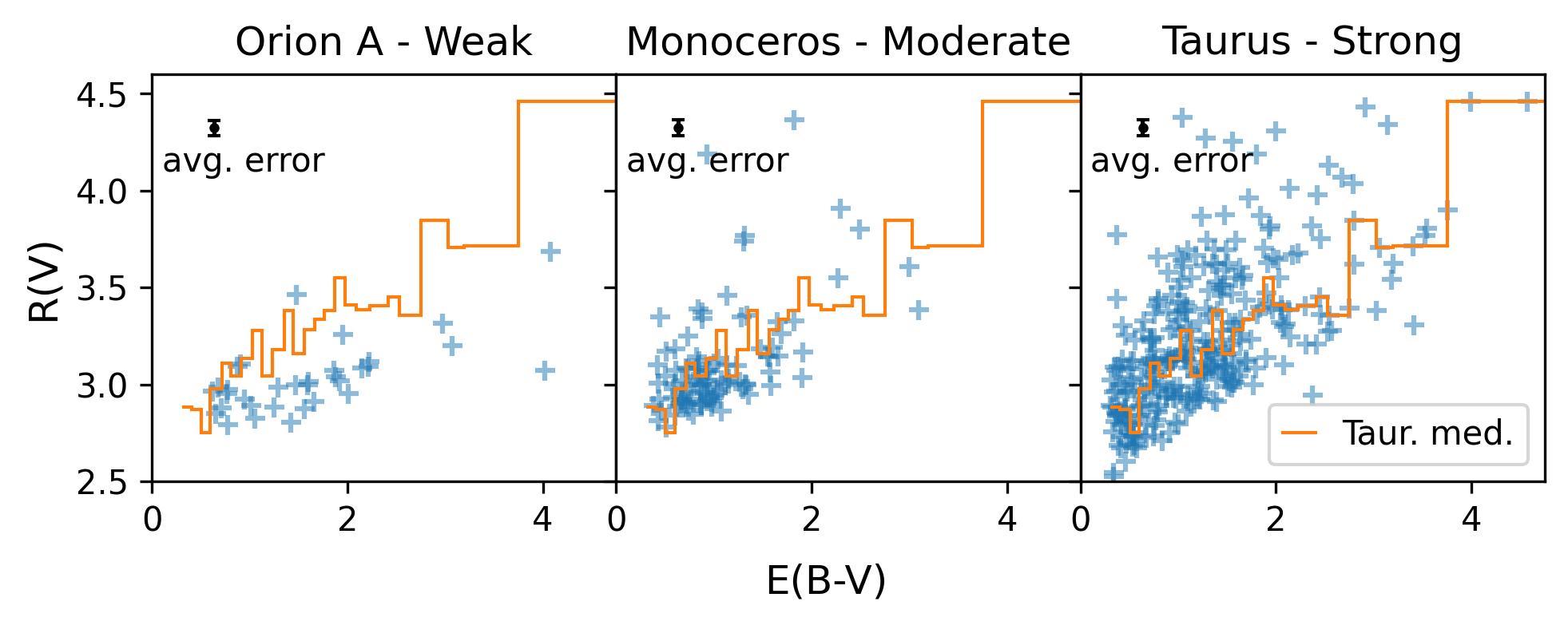}
        \caption{Example clouds for each category based on the correlation between \R\ and \E, where each + is a line of sight. The weak category is for clouds that do not exhibit much increase of \R\ with higher \E\ values. The moderate category is for clouds that show a moderate increase, but do not indicate well that \R\ would continue to increase at higher \E. Last is the strong category, which shows continual, strong \R\ increase with \E.\\
        We include a binned median  of Taurus \R s on each of the plots to help guide the eye. The Taurus median is binned by 0.1 along \E, unless there are less than 5 stars. If this is the case, then the bin widens until there are 5 stars within the boundary of the bin. This line shows that the correlation between \R\ and \E\ deviates across molecular clouds, as no clouds have the same relationship. The average error bars for the points in the clouds are shown in the top left of each panel, the method detailing how the errors are calculated is detailed in Appendix \ref{app:methods}.}
        \label{fig:cat}
    \end{figure*}
    
    \begin{figure*}
        \centering
        \includegraphics[scale=0.8]{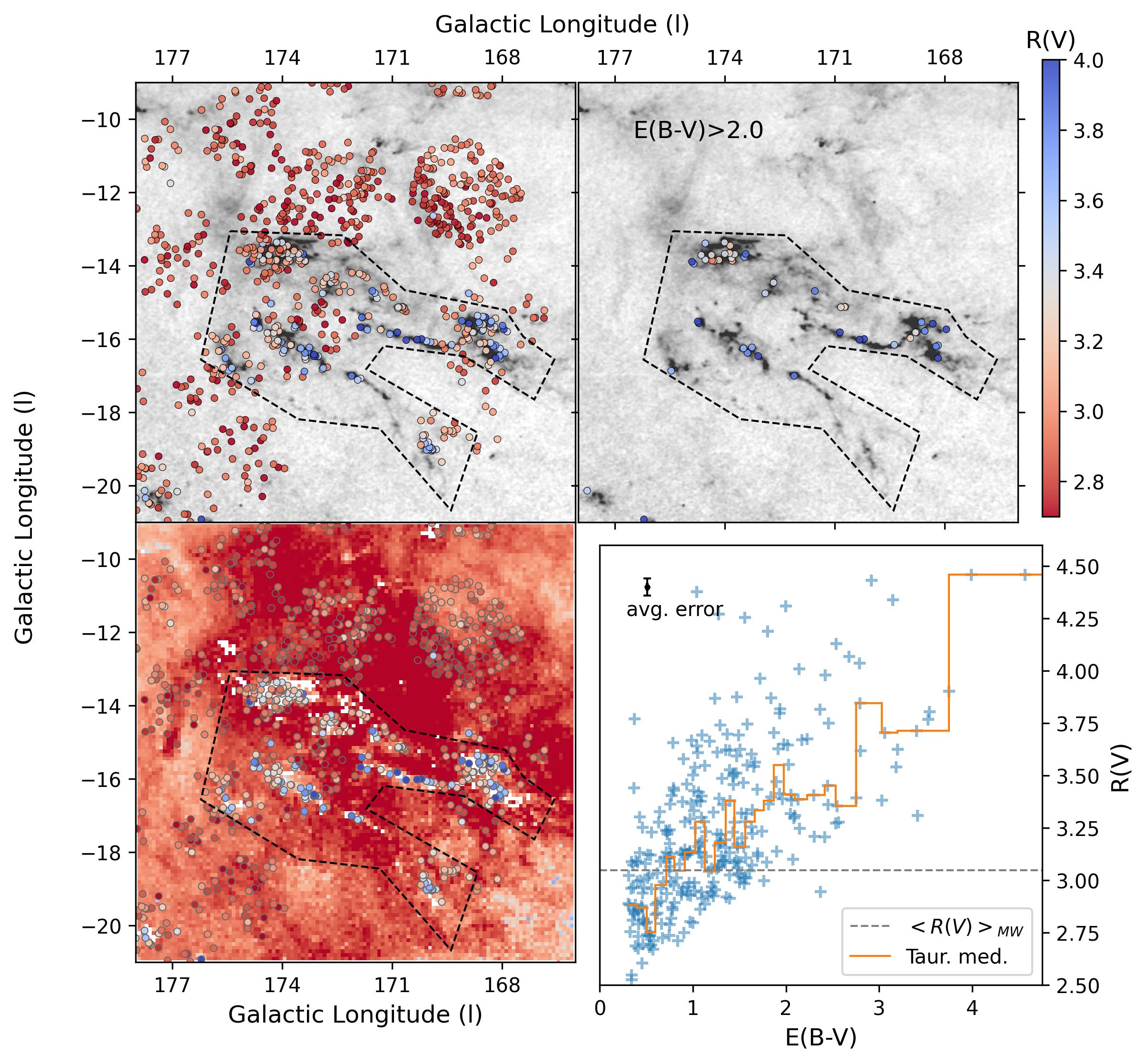}
        \caption{Extinction and \R\ in Taurus. The panels of the plots are as follows. \textbf{Top left}: line of sight \R\ values plotted as circles over the NICER/NICEST $A_J$ map. \textbf{Top right}: same as top left, except only points with $\E>2$ are plotted. \textbf{Bottom left}: line of sight \R\ values plotted as circles over the Z25 average \R\ map as a two dimensional histogram. The Z25 map's bins are $(0.1\degree)^2$, and any white spaces are NaN values, where measurements are missing. Our points often align with the white regions, demonstrating our work adds value to the all sky measurements. \textbf{Bottom right}: \R\ values plotted against \E\ to analyze if grain growth occurs in the cloud. The average \R\ of the Milky Way is shown in a dashed line, and a binned median of Taurus is shown in orange. This binned median trend is the same as that in Figure \ref{fig:cat}.}
        \label{fig:Taurus4}
    \end{figure*}
        
    \begin{figure*}
        \centering
        \includegraphics[scale=0.8]{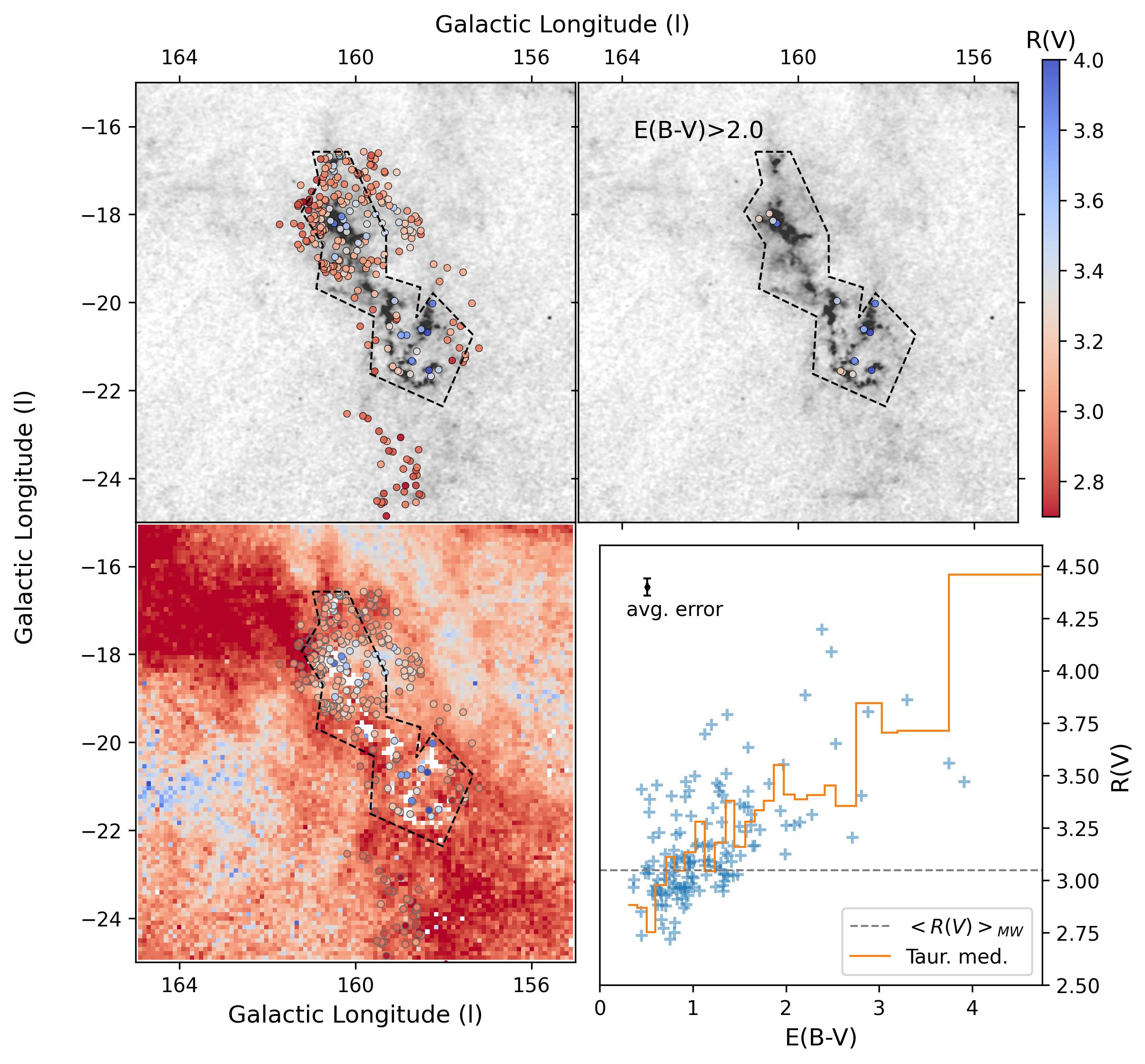}
        \caption{Same as Figure \ref{fig:Taurus4}, but for Perseus.}
        \label{fig:Perseus4}
    \end{figure*}

    \begin{figure*}
        \centering
        \includegraphics[scale=0.8]{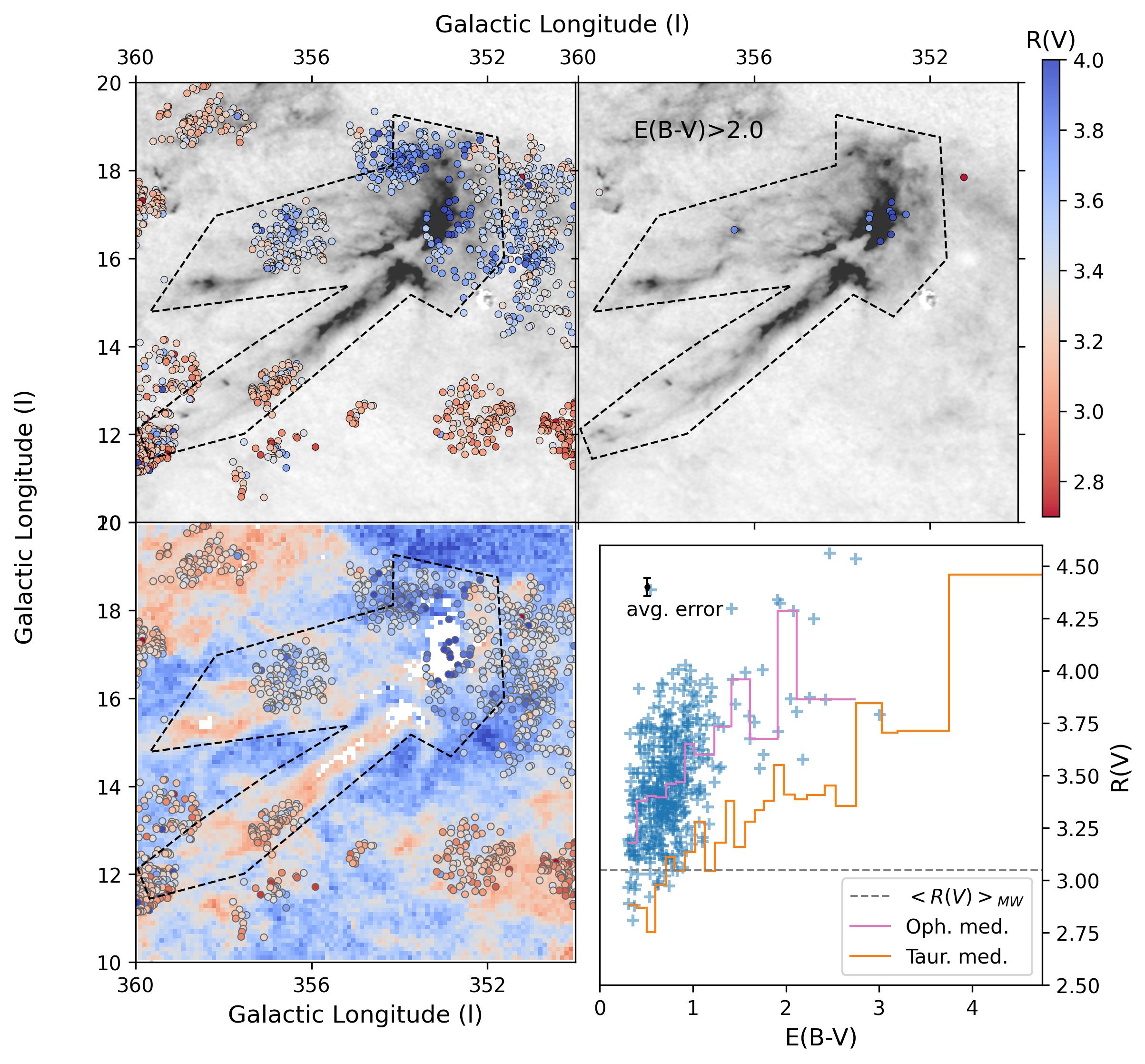}
        \caption{Same as Figure \ref{fig:Taurus4}, but for Ophiuchus. The additional line in panel 4, is the median \R\ for Ophiuchus which is binned by 0.1 along \E, unless there are less than 5 stars. If this is the case, then the bin widens until there are 5 stars within the boundary of the bin.}
        \label{fig:Ophiuchus4}
    \end{figure*}
    
    \begin{figure*}
        \centering
        \includegraphics[scale=0.8]{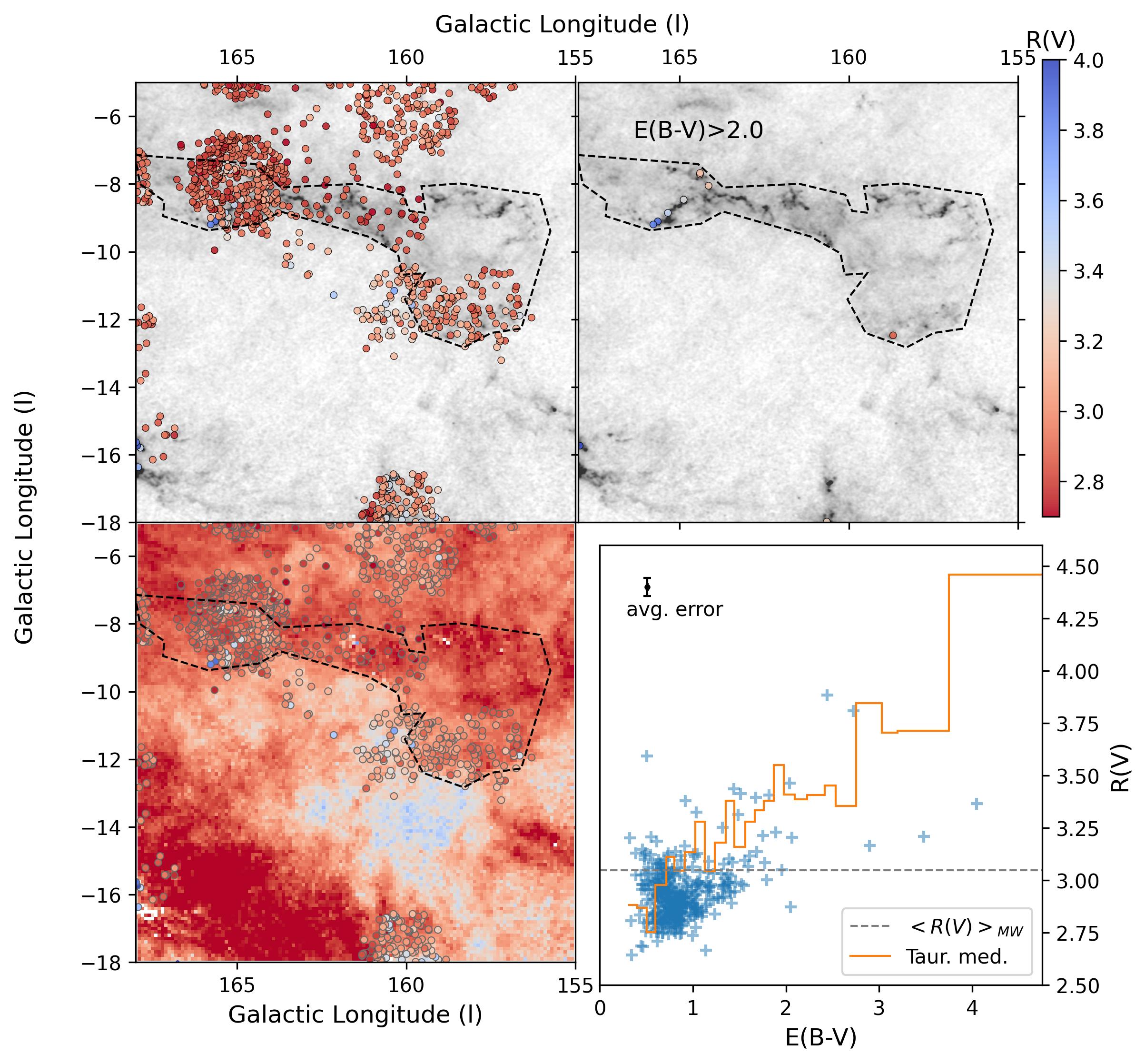}
        \caption{Same as Figure \ref{fig:Taurus4}, but for California.}
        \label{fig:California4}
    \end{figure*}
  
    \begin{figure*}
        \centering
        \includegraphics[scale=0.8]{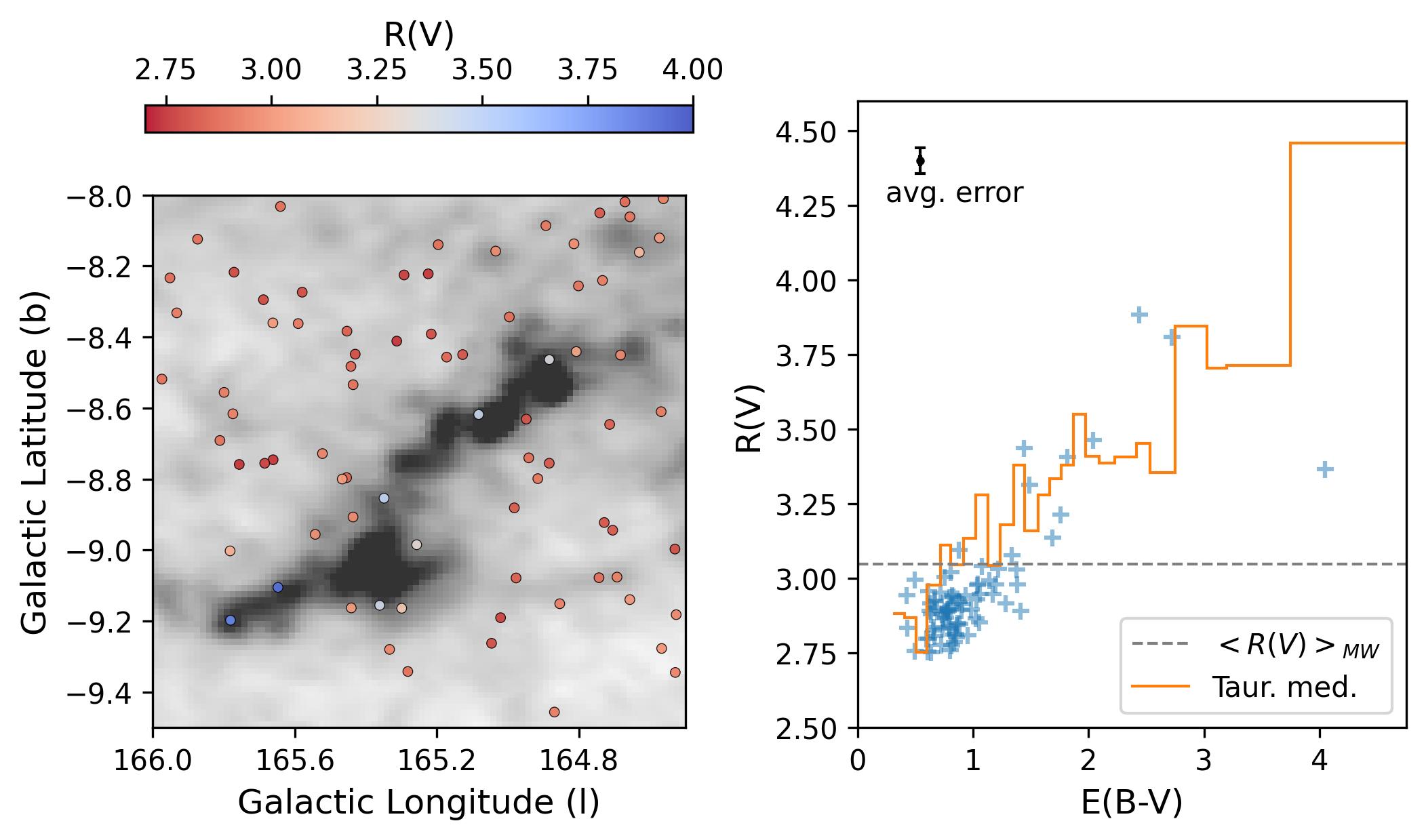}
        \caption{Same as the top left and bottom right plots in Figure \ref{fig:Taurus4}, but for NGC 1579.}
        \label{fig:NGC15794}
    \end{figure*}

    \subsection{Categorization}
    We find the relation between \R\ and \E\ varies significantly among nearby molecular clouds. To characterize this variation we divide the molecular clouds into three qualitative categories depending on the strength of the correlation: weak, moderate, and strong. Figure \ref{fig:cat} shows examples of the relationship between \R\ and \E\ for a representative cloud in each category. On these plots the median relationship derived from the Taurus molecular cloud is shown in orange to provide a common reference point. As evident in the Figure, clouds like Orion A show a relatively weak trend in \R\ with \E; there is little correlation until $\E\ \approx 2$, and \R\ is lower than the average value derived in Taurus for all but one line of sight. Meanwhile Monoceros, our example ``moderate'' cloud, shows a trend between \E\ and \R\ already at $\E = 1$. Finally, the Taurus molecular cloud shows an essentially linear relation between \E\ and \R\ starting at low columns ($\E \approx 0.5$), with very large scatter. These differences indicate the relationship between \R\ and \E\ is environmentally dependent, rather than universal.
    
    We provide our rough categorization of the different clouds in Table \ref{tab:clouds}. We further discuss the three clouds with the strongest trends---Taurus, Perseus, and Ophiuchus---as well as the California molecular cloud in detail in the following subsections. 
    Figures for the remaining clouds in our sample can be found in Appendix \ref{sec:appendix_4panels}.

    \subsection{Taurus}\label{subsec:Taurus_overview}
    Out of all of the clouds, Taurus has the strongest visual correlation between \R\ and \E. Figure \ref{fig:Taurus4} shows our coverage of Taurus on the NICER/NICEST A(J) map (upper left), the location of stars with the highest \E\ values on the NICER/NICEST A(J) map (upper right), our \R\ values on top of the Gaia \R\ map (lower left), and a scatter plot of \R\ versus \E\ (lower right). We show analogous plots for each cloud we discuss in detail (Sections \ref{subsec:Taurus_overview}-\ref{subsec:CA_NGC_Overview}). The upper left plot shows that we have stars sampling a variety of lines of sight across the cloud, from the dense filaments to the more diffuse outskirts. The upper right plot shows the largest values of \E\ and \R\ lie in these dense regions. The lower left plot shows that our measurements correlate well with the {\it Gaia}-derived \R\ measurements in the outskirts of the cloud, but we find a dramatic increase in \R\ as we move into the high extinction filaments that {\it Gaia} is not sensitive enough to probe. The lower right plot shows the relation between \E\ and \R\ observed in Taurus, starting at $\R \sim 2.8$ with $\E \sim 0.3$ ($A(V) \sim 0.8$) and increasing to $\R \sim 4.5$ for the handful of stars we observe with $\E \gtrsim 3.5$ ($A(V) \gtrsim 15.8$). The relation appears largely linear but with large dispersion; e.g., stars behind Taurus with $\E \approx 2$ can have \R\ as low as 3 ($A(V) \sim 6$) or as high as 4.25 ($A(V) \sim 8.5$). This result is consistent with previous work \citep{whittet1989,whittet_2001,mathis_1990} that showed Taurus had growing \R\, above the galactic average, and detected water-ice deposition in regions of higher extinction.

    \subsection{Perseus}
    The Perseus molecular cloud shows similar trends to those seen in Taurus, as illustrated in Figure \ref{fig:Perseus4}. As seen in the Figure, APOGEE spectroscopy does not sample Perseus and its surroundings as densely as it does the Taurus molecular cloud. We note however, there remains good coverage across the main body of the cloud (upper left), where \R\ ranges from about $3-4$. As in the case of Taurus, our measurements correlate neatly with the {\it Gaia} ones in the outskirts of the cloud (lower left), but we are able to probe to significantly higher \E\ and \R\ than {\it Gaia} using the deeper photometry and infrared sensitivity of our data sets. The trend between \E\ and \R\ (lower right) largely matches that seen in Taurus, with a strong correlation. This is consistent with results from \cite{mir_var_oph_per_2009}, which showed that towards higher column densities, a higher \R\ extinction curve is more consistent for these highly extinct regions.
    
    \subsection{Ophiuchus}
    The Ophiuchus molecular cloud has only patchy coverage with APOGEE spectroscopy, owing to the lack of APOGEE Reddening Survey targets in this cloud, as shown in Figure \ref{fig:Ophiuchus4}. However, we have a number of stars in the outskirts of the cloud and a handful deep in the main body of the cloud. Extinctions range up to $\E \lesssim 3$, and $3 < \R\ < 4.5$ (upper left). Our measurements confirm the {\it Gaia} measurements of \R\ in this cloud, which are substantially larger than in Taurus and Perseus (lower left). We find a strong trend between \E\ and \R, with a similar slope to the one found in Taurus, but shifted roughly by $0.4$ higher in \R\ at any given \E\ (lower right). This result is consistent with several previous studies that have noted the very high \R\ in the main body of Ophiuchus \citep{Oph_1980,mir_var_oph_per_2009,ir_var_oph_2023}. It should be noted that water-ice has only been observed at $A(V)\gtrsim13$ in this cloud \citep{whittet1980,tanaka1990}, while we observe increasing $R(V)$ with $A(V)$ starting at $A(V) \sim 1$, making it unlikely that this growth we observe is directly due to the formation of ice mantles.
    
    \begin{figure*}
        \centering
        \includegraphics[width=\linewidth]{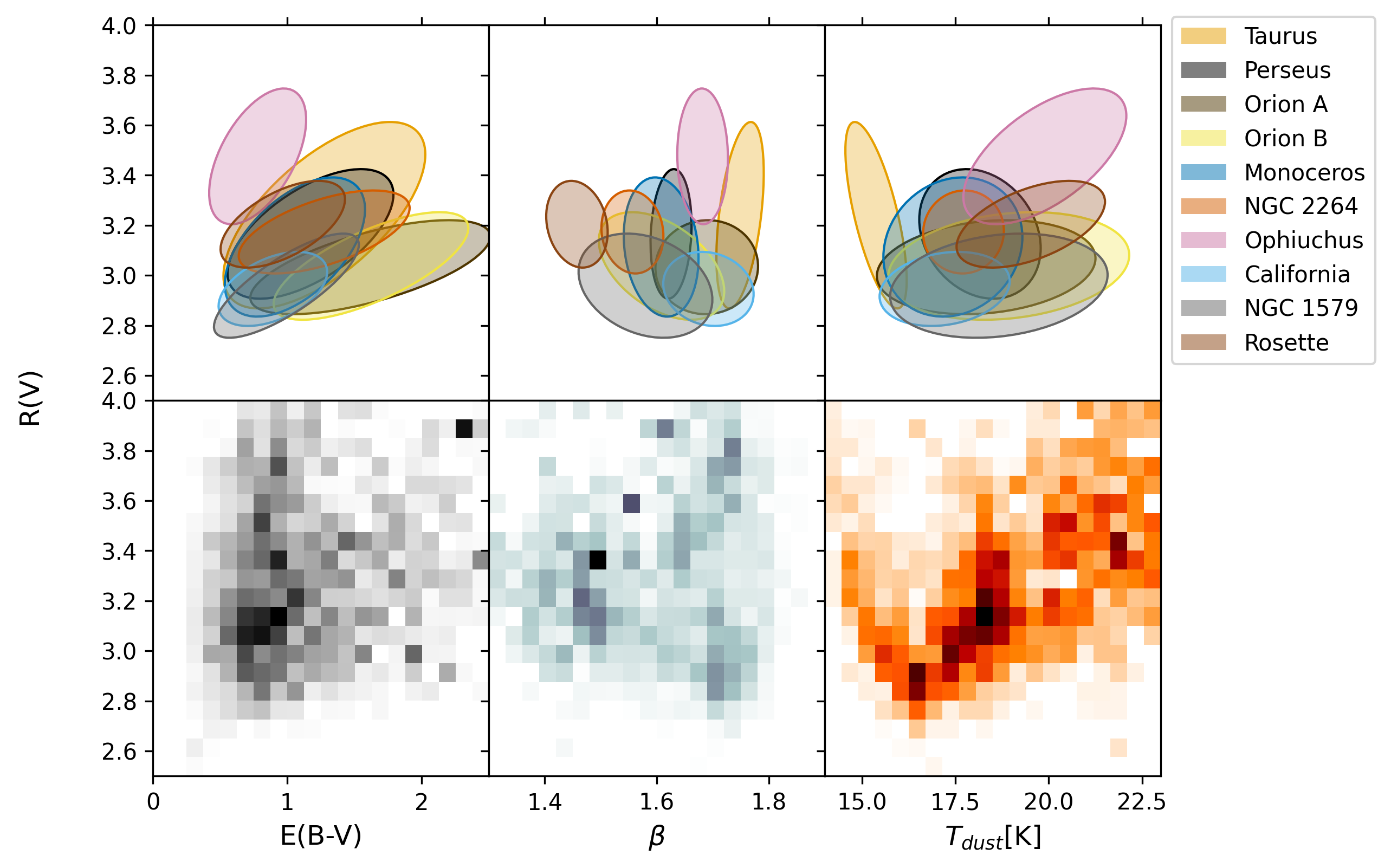}
        \caption{A comparison of the correlations of \R\ to different parameters associated with dust. \textbf{The top row} visualizes the correlation for each cloud through ellipses. Each ellipse is centered on the median values, scaled by one standard deviation in both directions, and aligned to the direction of the eigenvectors of the covariance matrix. \textbf{The bottom row} visualizes through a comparison of the entire sample rather than by individual clouds. The 2D histograms are weighted by the corresponding \textit{Planck} radiance to better distinguish which measurements are more representative of the dust.}
        \label{fig:all_corr}
    \end{figure*}

    \begin{figure*}
        \centering
        \includegraphics[width=0.83\linewidth]{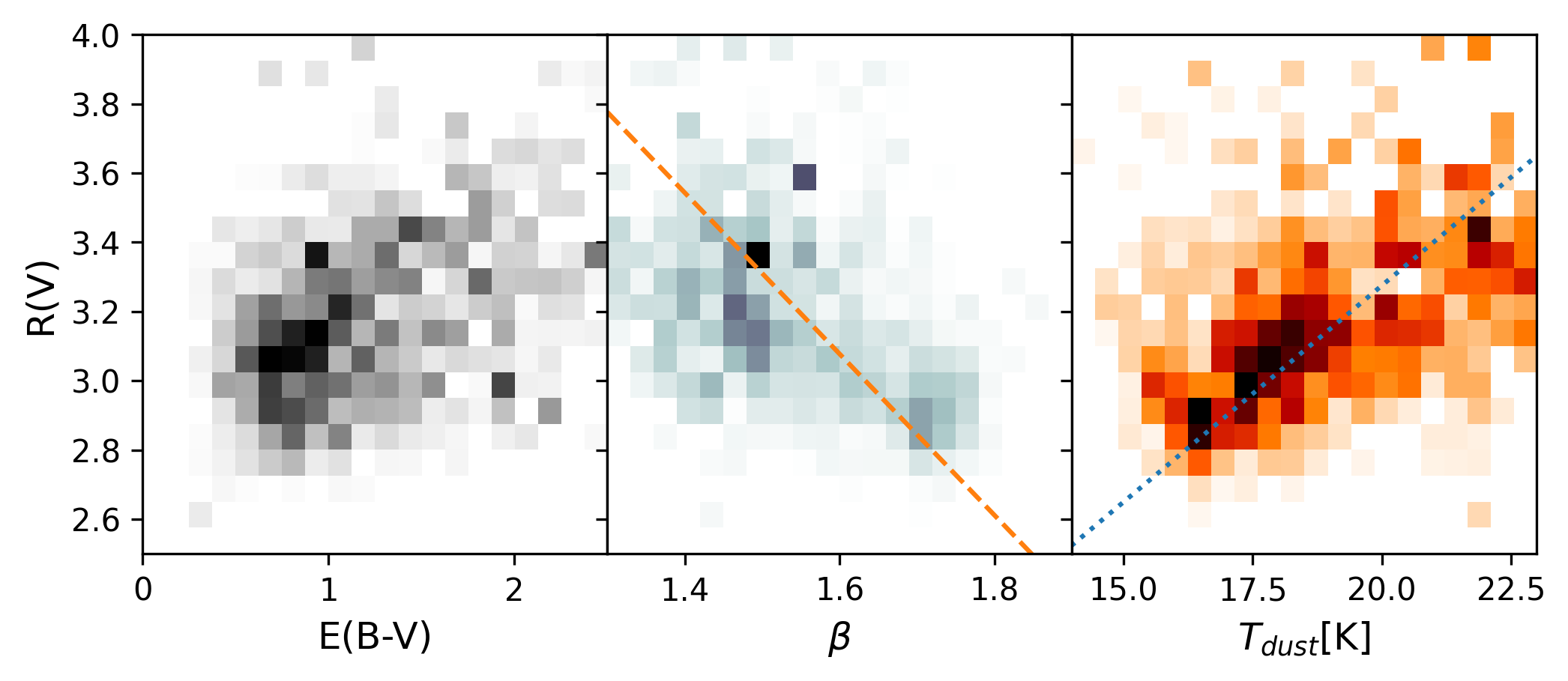}
        \caption{Same as the bottom row of Figure \ref{fig:all_corr}, but the stars for Taurus and Ophiuchus have been removed. The dashed orange line is a rough estimate of the linear relationship between \R\ and $\beta$ seen in S16, with a slope of -2.1. The dotted blue line is an estimate of the roughly linear relationship between \R\ and $T_\mathrm{dust}$, with a slope of 0.125.}
        \label{fig:maj_corr}
    \end{figure*}

    \subsection{California and NGC 1579}\label{subsec:CA_NGC_Overview}
    The California molecular cloud and the young cluster NGC 1579, located within California, show behavior rather different than the Taurus, Perseus, and Ophiuchus molecular clouds. As shown in Figure \ref{fig:California4}, our coverage of California ranges from excellent in the vicinity of NGC 1579 at (l, b) = (165.4, -9.0), to more scattered away from it (upper left plot). Across our measurements we find good correlations between our measurements and the {\it Gaia}-based \R\ measurements. Interestingly we find that outside NGC 1579, there is no apparent trend between \E\ and \R\ out to $\E \sim 4$. Meanwhile the handful of stars in the immediate vicinity of NGC 1579 show a strong relationship between \E\ and \R\ with remarkably low scatter, that falls almost exactly on the median trend we observe in Taurus, as shown in Figure~\ref{fig:NGC15794} in the right panel.

    The strong differences between the \E\ and \R\ trends in these two regions within the larger California molecular cloud indicates the strong role that properties beyond column density may play in shaping the grain size distribution. Clearly the vicinity of NGC 1579 is a rather atypical environment; this area is illuminated by the early B star LkH$\alpha$~101 \citep{lkhalpha}. Moreover, the fact that this part of the cloud hosts significant star formation and the rest does not suggests that independent of the massive star, this portion of California may be different from the other parts of the cloud. Our measurements make clear that \R\ is sensitive to these differences.

\section{Discussion}\label{sec:drivers}
    To investigate what drives the variation of \R\ in molecular clouds, we compare to various properties associated with dust. We inspect if \R\ is correlated with the proxy for dust column density (\E), dust spectral emissivity index ($\beta$), and dust equilibrium temperature ($T_\mathrm{dust}$).
    For these parameters we utilize the \textit{Planck} data release 1 all sky thermal dust model map\footnote{ HFI\_CompMap\_ThermalDustModel\_2048\_R1.20} to interpolate the values at the position of our background stars. \textit{Planck} calculated these values based on fitting the thermal emission of dust with a modulated blackbody distribution. We choose to use the first data release as it includes the fewest assumptions in the data product.
    
    The following sections are split into investigating trends across the entire molecular cloud sample, and individually examining clouds with the strongest $R(V)-\E$ correlation, Taurus and Ophiuchus. For these individual clouds, we search for regional correlations within each cloud, motivated by our observations of the different behavior of \R\ in the vicinity of NGC 1579 in California.
    
    \begin{figure*}
        \centering
        \includegraphics[scale=0.85]{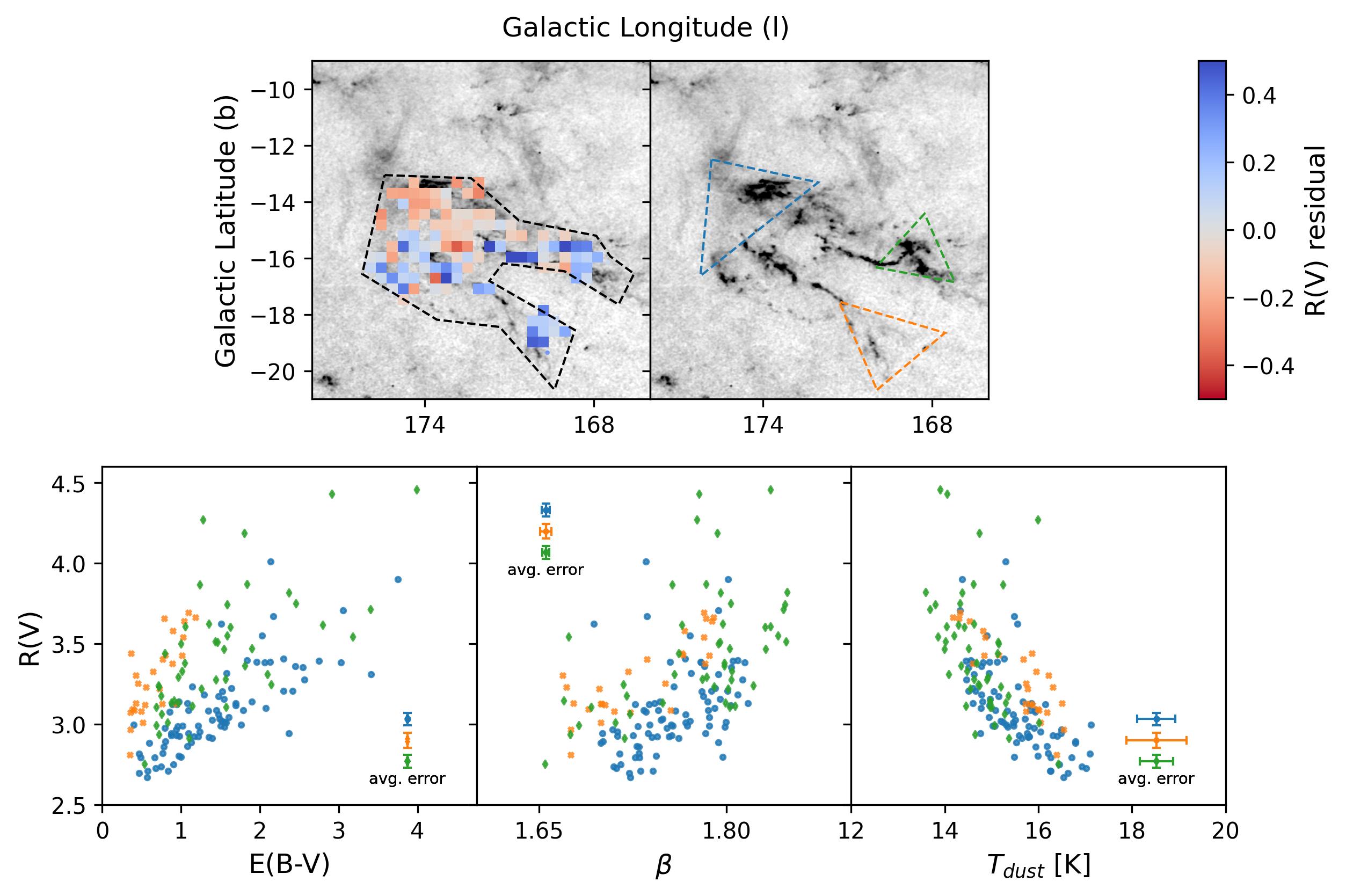}
        \caption{In the \textbf{top row}, the left plot shows the spatial map of the residual \R\ values for Taurus, and the right plot displays the sub-regions we made based on the residual map. The plots in the \textbf{bottom row} show the relationship between \R\ and other properties where points have been colored by the region with the corresponding color on the spatial map, with median errors shown for reference.}
        \label{fig:Taurus_reg}
    \end{figure*}

    \subsection{Relationships Across the Sample}\label{sec:rel}
    We compare the trends of \R\ with \E, $\beta$, and $T_\mathrm{dust}$ across our entire sample in Figure \ref{fig:all_corr}. In order to make both between-cloud and within-cloud variations salient, for each cloud in the top row of Figure~\ref{fig:all_corr} we show a single ellipse, derived from the measured dispersions and correlation of the points in each cloud. The ellipses are centered on the median values, scaled by one standard deviation, and oriented along the eigenvectors of the covariance matrix. In the bottom row, we show the entire sample in a 2D histogram weighted by the corresponding \textit{Planck} radiance to more strongly feature measurements with high dust emission. Here we can see that \R\ and \E\ have a general positive correlation across clouds (top left panel), albeit with significant variation between clouds; meanwhile correlations between \R\ and $\beta$ or $T_\mathrm{dust}$ are harder to make out (right two columns). 
    
    The plot of \R\ versus $T_\mathrm{dust}$ (bottom right) shows a peculiar trend where at low temperatures $T_\mathrm{dust}$ and \R\ are negatively correlated, while at higher temperatures they are positively correlated. This behavior is driven largely by the Taurus molecular cloud. Both it and Ophiuchus show significantly different distributions than the other seven clouds we consider in this work. We explore removing those two clouds in Figure \ref{fig:maj_corr}. In the remaining clouds, \R\ shows clear correlations with \E, $\beta$, and $T_\mathrm{dust}$, with significantly reduced dispersion. The dashed orange line (middle panel) shows how our results compare with those seen in S16; we find decent agreement, though our data show a somewhat shallower slope.

    Taurus and Ophiuchus have very different trends than the other clouds we consider. Ophiuchus generally has higher \R\ values than the rest of the sample, in a way that cannot be explained by low temperature or low $\beta$. That said, the slope of the relation between \R--$T_\mathrm{dust}$ is similar to other clouds. Taurus is another outlier in our sample. We find that the extinction curve flattens as $\beta$ steepens in Taurus, which is the opposite trend to what we find elsewhere. This is consistent with the results of \cite{planck_mc, planck_ant_cent_mc}, which saw an anti-correlation between $\beta$ and $T_\mathrm{dust}$ alongside lower temperature and higher $\beta$ values in dense regions of Taurus.
    
    We further discuss the implications of these different trends in \textsection\ref{subsec:RvT}. We now turn to analyzing variations in these outlier clouds, aiming to uncover meaningful differences like that between NGC 1579 and the rest of California.

    \begin{figure*}
        \centering
        \includegraphics[scale=0.85]{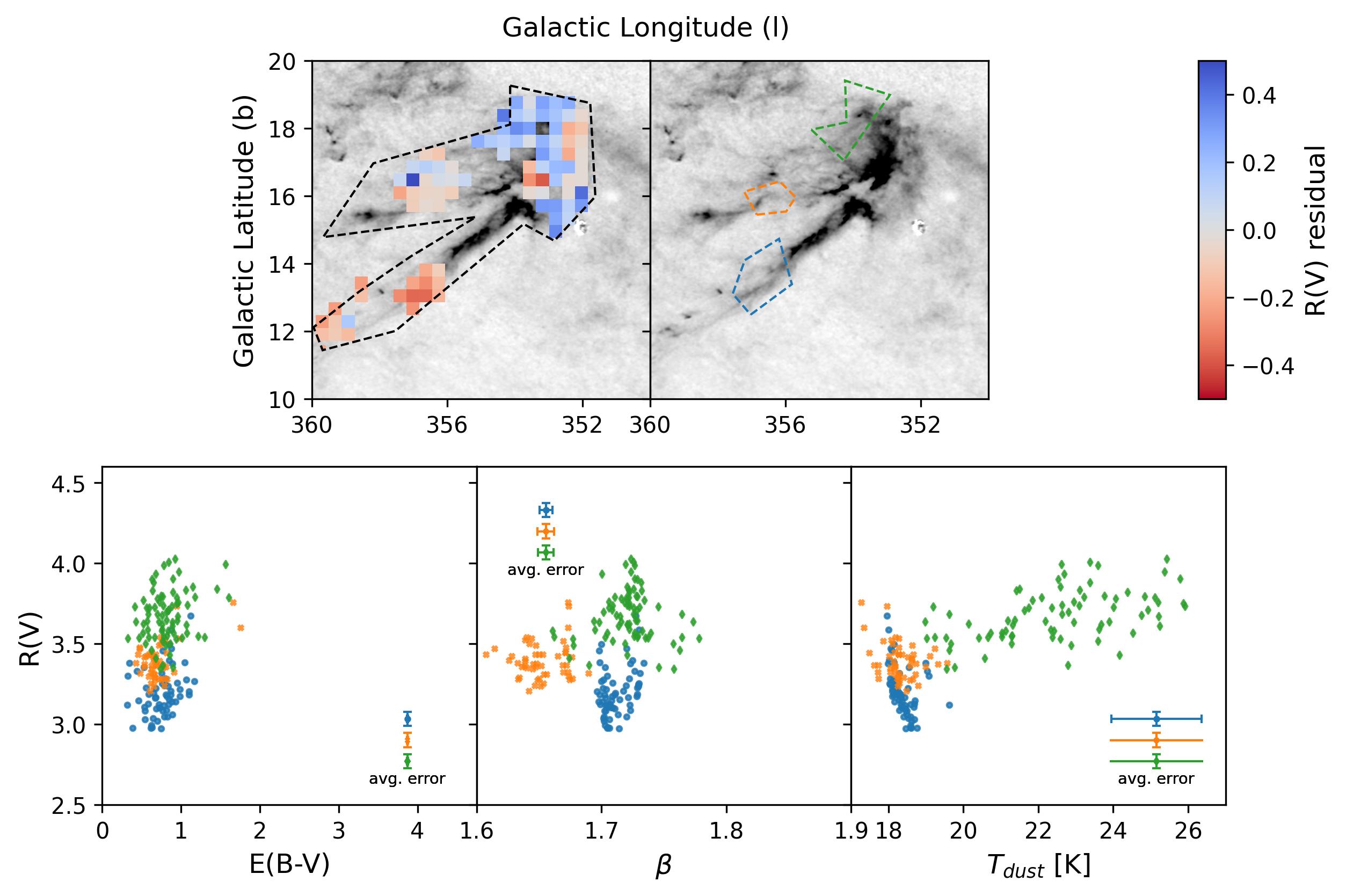}
        \caption{Same as Figure \ref{fig:Taurus_reg}, but for Ophiuchus.}
        \label{fig:Ophiuchus_reg}
    \end{figure*}
    
    \subsection{Regional Variation}
    We look for variations in the trend between \R\ and \E\ for our distinctly different clouds, Taurus and Ophiuchus. Both of these clouds show far more variation in \R\ with \E\ than can be explained by the statistical uncertainties alone. We explore the spatial coherence of these variations by performing a linear fit through the binned median values (as seen in the bottom right of Figure \ref{fig:Taurus4}), and spatially mapping the residuals ($R(V)_{resid} = \R-\R_{med}$) over the cloud. We then inspect and divide the residual maps, by eye, into separate regions, and plot \R\ against different probes of the interstellar medium for each region. The maps for the other clouds can be found in Appendix \ref{sec:resid_maps}. We discuss the results for Taurus and Ophiuchus in detail here.
    
    \subsubsection{Taurus}
    We show the residual spatial map for Taurus in Figure \ref{fig:Taurus_reg} in the top left panel. Blue regions show areas where the typical \R\ values are above expectations from the linear fit, while red regions are below. There is a clear relationship between the residual \R\ and the location within the cloud. The northern region of Taurus falls below the linear fit, while the southern regions fall above the linear fit. We use this map to divide the cloud into regions for comparison, shown in Figure \ref{fig:Taurus_reg} in the upper right panel.

   The top right panel of Figure \ref{fig:Taurus_reg} shows the regions we choose, and how \R\ depends on different properties within each region. Strikingly, within each region of the cloud the scatter in the trend between \E\ and \R\ is greatly reduced; much of the scatter seems to be driven by relatively large-scale differences in the properties of the cloud in different regions rather than by smaller scale differences. We do not see this same separation in the correlations of \R\ with $\beta$, and $T_\mathrm{dust}$.
   
    \subsubsection{Ophiuchus}
    We map the residuals to a simple linear fit of \R\ to \E\ in Ophiuchus in Figure~\ref{fig:Ophiuchus_reg}. As in Taurus, the fit residuals show substantial spatial correlations. The limited spatial coverage of the cloud in the APOGEE spectroscopy somewhat limits our ability to cut the cloud into different regions, but we select three different regions with good APOGEE coverage (Figure~\ref{fig:Ophiuchus_reg}, top right panel).

    These three regions show significant differences in \R, but the range of \E\ probed in each is too limited to make good measurements of the respective slope of \R\ with \E. Similarly the regions have different $\beta$, and $T_\mathrm{dust}$. A moderate correlation between \R\ and $T_\mathrm{dust}$ is apparent when combining the different regions---also visible in Figure~\ref{fig:all_corr} for Ophiuchus---but within a given region no correlation is apparent.

\subsection{Implications for estimating the total mass of molecular clouds}

The work of \cite{Bohlin_1978} is frequently used in studies of molecular clouds to convert from measured extinctions to total masses of hydrogen. They derive a mean ratio of total neutral hydrogen to color excess of 
\begin{equation}
\langle N(H_I + H_2)/\E\rangle = 5.8 \times 10^{21} \text{atoms} ~\text{cm}^{-2}~\text{mag}^{-1} \, .
\label{eq:bohlin}
\end{equation}
For these conversions it is often assumed that $\R=3.1$, allowing $A(V)$ to be used instead of \E. However, variation in the extinction curve will clearly change the relationship between hydrogen column density and the extinction in the optical. Accordingly, we consider here what level of effect one might expect our observed variation in $R(V)$ to have on computed molecular cloud masses from Equation~\ref{eq:bohlin}.

We measure \R\ in different molecular clouds from a low of 2.9 in California to a high of $\sim 4.5$ in Taurus. This implies a roughly 50\% peak-to-peak uncertainty in estimating molecular cloud masses, if one has no reason to prefer \E\ or $A(V)$ as a proxy for amount of molecular material. Practically, however, one expects that most of the grain mass would be in large grains, and so we expect extinction curve measurements that are less sensitive to the number of small dust grains to show lower variation in $N(H) / A$. 

To make a more conservative estimate for the variation in $N(H) / A$, we consider how the ratio 
\begin{equation}
    \frac{A(K) - A(W2)}{A(J) - A(K)}
    \label{eq:redratio}
\end{equation} changes with \R. The numerator of Equation~\ref{eq:redratio} is a rough estimate for the K band extinction, based on comparison with redder W2 photometry, while the denominator is the familiar $J-K$ color excess used for estimating molecular cloud masses in works like \citet{Juvela_NICERST}. According to the extinction curve of \citet{Schlafly_2016}, when \R\ changes by 30\% from 3.3 to 4.3, the value of Equation~\ref{eq:redratio} changes by 22\%---variations in the extinction curve are indeed smaller in the infrared than in the optical, but not hugely so. We conclude that variation in \R\ can induce uncertainties in mass at up to the 30\% level. However, this represents the most extreme variation possible across our sample of molecular clouds, comparing the highest and lowest \R\ we measure. We estimate a more typical mass uncertainty by measuring the scatter in $\langle \R \rangle$ among our nine molecular clouds, as tabulated in Table~\ref{tab:clouds}, which gives  $\sigma_{\R} ~ \sim0.15$, corresponding to a typical implied mass uncertainty of less than $\sim5\%$.

The preceding discussion has focused on the largest variations we see among all of the stars in our sample and the typical dispersion we see on average between different molecular clouds. An additional important consideration in translating $A(V)$ to mass is that we find that $R(V)$ depends on $A(V)$, so adoption of a single conversion between $A(V)$ and $N(H)$ for a molecular cloud leads to systematic errors in the dense regions of clouds relative to their outskirts. The naive expectation is that grains are growing, so $N(H) / A(V)$ should increase with $R(V)$, leading fixed $R(V)$ prescriptions to underestimate the mass of dense regions. This makes a small effect on the overall mass of the cloud, since $R(V)$ variations are limited in low extinction regions. For example, we find that variations in \E\ are greatest when $\E \gtrsim 2$, but the work of \citet{lombardi_2011} suggests that only 10\% of the mass of a molecular cloud resides at those column densities. Nevertheless, it is precisely these high density regions that are the most relevant for star formation.

\subsection{$R(V)$, ice mantle formation, and $T_\mathrm{dust}$}\label{subsec:RvT}
In Section \ref{sec:rel}, we found that most clouds had a positive correlation between \R\ and $T_\mathrm{dust}$, while Taurus had a negative correlation.  Here, we discuss the interpretations and implications of this result.

Among the clouds we look at, the trend in \R\ with $T_\mathrm{dust}$ in Taurus is most easily interpreted in the context of the formation of ice mantles in dense, shielded molecular clouds. We find that $R(V)$ increases with decreasing temperature (and therefore increasing shielding) in Taurus. Meanwhile, literature results find that ice mantle formation sets in at $A(V)\sim3$ \citep{whittet1989,whittet_2001,whittet2007} in this cloud. We find $R(V)$ increasing with $A(V)$ starting at $A(V) \lesssim 1$ for this cloud, which is lower than, but similar to, the threshold for ice formation. So our observations of increasing $R(V)$ with $E(B-V)$ and decreasing temperature fit reasonably well with a picture for grain growth due to ice mantle formation (Figure \ref{fig:all_corr}).

Ice mantle formation has also been observed in Ophiuchus, starting at a very high threshold of $A(V)\sim13$ \citep{whittet1980,tanaka1990}. This is roughly the highest $A(V)$ in our sample, but we observe $R(V)$ increasing with $E(B-V)$ in Ophiuchus down to the roughly the lowest $A(V)$ we find reliable, $A(V) \lesssim 1$. Moreover, in Ophiuchus we find that $R(V)$ tends to increase, rather than decrease with temperature. So our measurements of increasing $R(V)$ with $A(V)$ in Ophiuchus do not fit cleanly into an ice-mantle formation picture and would instead need to be driven by effects like coagulation.  

The high threshold for ice formation in Ophiuchus is usually attributed to the higher ambient radiation field in that region \citep{tanaka1990}, which inhibits ice formation due to photo-desorption. In Ophiuchus we find a higher average $R(V)$ but a similar increasing trend between $R(V)$ and $E(B-V)$ as in Taurus. We conclude that the radiation field does not inhibit grain growth in the same way that it inhibits ice mantle formation.

Outside Taurus and Ophiuchus, we find a good, positive correlation between \R\ and $T_\mathrm{dust}$ (Figure~\ref{fig:maj_corr}), with $\delta \R \approx 0.125 \cdot \delta T_\mathrm{dust}$. This correlation is broadly consistent with a picture where small grains are more easily destroyed in diffuse regions of molecular clouds with limited shielding \citep{Zhang_2025,lightcube}. In this picture, higher temperatures imply fewer small grains and a flatter extinction curve. Dust temperature is closely tied to the interstellar radiation field density $u$.

The work of \cite{lightcube} finds a positive correlation between $u$ and \R, at least for low $u$. For high dust-density regions, they find roughly
\begin{equation}
\delta \R \approx \frac{2}{3}\,\delta \log_{10}(u) \, ,
\end{equation}
which we can compare with our derived $R(V)$--$T_\mathrm{dust}$ relationship. Assuming $T\propto u^{1/6}$ and a reference temperature of 20K, we derive $\delta \R \approx 0.087 \cdot \delta T_\mathrm{dust}$, compared to our $\delta \R \approx 0.125 \cdot \delta T_\mathrm{dust}$. This agreement is rather good given the dispersion in our observed relationship and the different behaviors \cite{lightcube} finds in low and high density regimes. The completely independent measurements (different sources of \R, $T_\mathrm{dust}$, and portions of the sky) suggests that the positive $R(V)$--$T_\mathrm{dust}$ is a common feature of the interstellar medium.

This picture has the awkward consequence that it predicts high $R(V)$ for the diffuse interstellar medium. The typical temperature of the high Galactic latitude ISM is $\sim21.5$K \citep{Planck_2014}. According to the relation in Figure \ref{fig:maj_corr}, these diffuse environments would have a $\langle \R \rangle\sim3.4$, rather higher than the traditional value of 3.1. On the other hand, our fields are specifically selected to probe molecular clouds, and are therefore rather different from typical diffuse ISM sight lines. We would require additional environmental dependence alongside temperature to be able to explain the correlation between \R\ and $T_\mathrm{dust}$ in most of our molecular clouds, unlike the diffuse ISM.


\subsection{Other Potential Causes of Variation in \R}\label{subsec:causes}
We find significant variation in the relationship between \R\ and other ISM dust tracers across our sample. This variation could be caused by additional local, physical properties we are not accounting for. We discuss here possible environmental effects that could differ between clouds to change their \R--\E\ relationships.

The work of \cite{acc_coag_2012} explores how accretion, coagulation, and turbulence interplay in simulations. The study finds that grains are only able to grow to very large sizes through turbulence after the gas-phase metals have been used in accretion. Although they find that coagulation flattens the features of the extinction curve, it is a very small suppression. Assuming the standard dust-to-gas ratio \citep{Bohlin_1978} and a typical cloud thickness of 15 pc, we estimate that their simulations reach $\E \sim 2.4$ ($A(V) \sim 7.5$). The large variance we observe in the \R--\E\ relationship in molecular clouds spans both weak flattening consistent with their predictions and stronger flattening not reproduced by their simulations. They note that this small suppression is inconsistent with observations and would require very strong coagulation in order to reproduce similarly flattened extinction curves.

\cite{acc_coag_2012} discuss several mechanisms that may help reconcile these models with observations. If turbulence drives grain growth, then clouds with the highest \R\ values may preferentially correspond to regions with large velocity dispersions. Additionally, strong coagulation may occur in extremely dense regions, such as those with star formation. This matches our observations of the California cloud, where only the star-forming sub-region of NGC 1579 has a strong \R--\E\ relationship. Furthermore, longer cloud lifetimes could lead to a higher abundance of larger grains. Although the age of a cloud is very challenging to measure, it could vary dramatically across clouds and lead to differences in grain size distributions through the timescales available for accretion and coagulation.

Alongside local conditions of the clouds, the inherent diversity of grain species could be a large contributor to the \R\ variations. In \cite{themis_2_2024} they explore how models of differing grain species' shape, mantle, and base chemical composition affect the efficiency of absorption, scattering, and polarization. They find wavelength-dependent absorption/scattering efficiencies, and that the strength of those relationships changes with mantle thickness. The large level of dispersion in their models due to differing grain chemistry may contribute to much of the dispersion we observe in our molecular clouds. This could affect both internal cloud dispersion as well as the inhomogeneity across clouds, due to chemical differences on small scales and in different cloud environments. However, it is unclear to us how significantly different chemical compositions would be obtained in different molecular clouds.

Outside of the properties of the grains, the inherent geometry of molecular clouds may play a significant role in the observed variation. Molecular clouds have complex geometries, composed of sheets, filaments, and dense cores of material \citep{mc_sheets,mc_filaments,Herschel_2010}. These structures create a complex path for radiation to traverse, resulting in a varying level of permeating radiation depending on the location within the cloud. In turn, the projection of the cloud mixes structures along the line of sight, hiding pockets of higher density that can facilitate grain growth. As the geometry varies between clouds, this affects how we interpret the local environments and may produce variations in \R\ through differences in density along the line of sight.

Taken together, these results suggest that additional local properties may contribute to the \R\ variations in molecular clouds. The observed variation arise from environmental evolution, intrinsic grain chemical diversity, or the geometry of the clouds themselves. Distinguishing which of these factors, if any, drives the dispersion we observe will require further work.

\section{Conclusion}\label{sec:conclusion}
We have measured the shape of the extinction curve as parameterized by \R\ for 52,912 stars above $\delta = -30\degree$, enabled by APOGEE spectroscopy and PS1, 2MASS, and WISE photometry. Our catalog is publicly available on \href{https://zenodo.org/records/22117631}{zenodo}. Using these measurements, we analyze how the extinction curve varies in nearby molecular clouds. We make the following conclusions:

\begin{itemize}
    \item Most, but not all, molecular clouds show at least some positive correlation between \R\ and \E, consistent with the expectation that grains grow in dense regions in molecular clouds. This trend was not seen in S16, likely because their analysis focused on Galactic-wide variations rather than nearby dense molecular sight lines.
    
    \item The trend between \E\ and \R\ varies significantly from cloud to cloud. Clouds like Taurus show a strong relationship between \E\ and \R, while Orion A shows little correlation.
    
    \item After removing Taurus and Ophiuchus, which appear as outliers, we find a negative correlation between $\beta$ and \R, and positive correlation between $T_\mathrm{dust}$ and \R. This is broadly consistent with small grains being more easily destroyed in diffuse regions of molecular clouds with limited shielding. However, if this relationship were to hold generically in the diffuse ISM, it would imply larger $\langle \R \rangle$ than measured, complicating this picture.

    \item In the Taurus molecular cloud, we find a significant negative correlation between $T_\mathrm{dust}$ and \R. This is consistent with grain growth in dense, shielded regions of molecular clouds through the formation of ice mantles, but surprisingly we only find this correlation in Taurus.
    
    \item Local environments within a cloud can influence the relationship between \E\ and \R. In Taurus, our best-sampled cloud, the trend between \E\ and \R\ tightens dramatically if we split the cloud into several smaller regions. However, we were not able to determine what drove these observed environmental variations.
    
    \item Local conditions associated with star formation can dramatically affect the $\R-\E$ relationship. In the California Molecular Cloud, we see a strong positive correlation much like that observed in Taurus along lines of sight in the immediate vicinity of the NGC 1579 young stellar cluster. Outside that region, we see essentially no correlation between \E\ and \R.
    
\end{itemize}

Although we do not know the exact reason these differences arise, we understand that it likely reflects local conditions unique to each cloud. Further work is needed in order to understand what properties drive the \R\ variation in these dense environments.

Some possible drivers that we do not explore in this work are age, cloud geometry, grain chemistry, and turbulence. Our results motivate further investigation into the physics of dust grains within molecular clouds. Identifying the drivers of the extinction curve will require deeper observations and larger samples, which will be enabled with several upcoming surveys such as LSST, Roman, and SDSS-V. Understanding how dust properties vary in dense environments is essential for accurately interpreting extinction throughout the Milky Way and other galaxies.

\begin{center}
    \textit{Acknowledgments}
\end{center}

\textit{Software}: \texttt{astropy} \citep{astropy2013,astropy2018,astropy2022}, \texttt{SAO DS9} \citep{DS9}

We thank the anonymous referee for important feedback that significantly improved this work.

MH-R acknowledges support from the STScI DRF.

This work made use of ChatGPT, a large language model developed by OpenAI, to assist with small amounts of coding, debugging, and draft editing. ChatGPT was accessed via the ChatGPT web interface (https://chat.openai.com).

The Pan-STARRS1 Surveys (PS1) and the PS1 public science archive have been made possible through contributions by the Institute for Astronomy, the University of Hawaii, the Pan-STARRS Project Office, the Max-Planck Society and its participating institutes, the Max Planck Institute for Astronomy, Heidelberg and the Max Planck Institute for Extraterrestrial Physics, Garching, The Johns Hopkins University, Durham University, the University of Edinburgh, the Queen's University Belfast, the Harvard-Smithsonian Center for Astrophysics, the Las Cumbres Observatory Global Telescope Network Incorporated, the National Central University of Taiwan, the Space Telescope Science Institute, the National Aeronautics and Space Administration under Grant No. NNX08AR22G issued through the Planetary Science Division of the NASA Science Mission Directorate, the National Science Foundation Grant No. AST-1238877, the University of Maryland, Eotvos Lorand University (ELTE), the Los Alamos National Laboratory, and the Gordon and Betty Moore Foundation.

This publication makes use of data products from the {\it Two Micron All Sky Survey}, which is a joint project of the University of Massachusetts and the Infrared Processing and Analysis Center/California Institute of Technology, funded by the National Aeronautics and Space Administration and the National Science Foundation.

This publication makes use of data products from the {\it Wide-field Infrared Survey Explorer}, which is a joint project of the University of California, Los Angeles, and the Jet Propulsion Laboratory/California Institute of Technology, funded by the National Aeronautics and Space Administration.

This work has made use of data from the European Space Agency (ESA) mission {\it Gaia} (\url{https://www.cosmos.esa.int/gaia}), processed by the {\it Gaia} Data Processing and Analysis Consortium (DPAC, \url{https://www.cosmos.esa.int/web/gaia/dpac/consortium}). Funding for the DPAC has been provided by national institutions, in particular the institutions participating in the {\it Gaia} Multilateral Agreement.

Based on observations obtained with {\it Planck} (\url{http://www.esa.int/Planck}), an ESA science mission with instruments and contributions directly funded by ESA Member States, NASA, and Canada.

\bibliography{main}{}
\bibliographystyle{aasjournalv7}

\appendix

\section{Obtaining parameters and uncertainties}\label{app:methods}
    To solve for $E'$ and x, we define the diagonal matrix \textbf{C}, with the squared photometric errors in each band:
            $$\textbf{C} =
              \begin{bmatrix}
                \sigma_{g}^2 & & \\
                & \ddots & \\
                & & \sigma_{W2}^2
              \end{bmatrix}.$$
        This matrix weights each band according to its uncertainty, and allows us to calculate the covariance of \textbf{y}:
            $$Cov_y=(\textbf{A}^T\textbf{C}^{-1}\textbf{A})^{-1}.$$
        The best-fit parameters and corresponding uncertainties are then given by:\\
            $$\textbf{y}=\mu, E', E'x = (Cov_y \textbf{A}^T \textbf{C}^{-1})(\textbf{m}_{obs} - \textbf{m}_{int}),$$
            $$\sigma_\mu, \sigma_{E'}, \sigma_{E'x} = (Diag(Cov_y))^{1/2}.$$
        
        Having obtained the \textbf{y} components, we propagate the uncertainties to compute the error of \R:
            $$\sigma_{R(V)} = 9.1 * \sqrt{(\frac{\sigma_{E'x}}{E'})^2 + (\frac{x\sigma_{E'}}{E'})^2}.$$
        This procedure, paired with equations \ref{eq_E} and \ref{eq_RV}, obtains proxies for \E, \R, and corresponding uncertainties for all stars in our sample.

\section{\R\ and \E\ relationship for all clouds}\label{sec:appendix_4panels}
    In the main body of the paper we limited our individual cloud analysis of the \R\ and \E\ relationship to clouds with the strongest correlations. Here we provide figures of the remaining clouds in our catalog.
    
    \begin{figure*}[h!]
        \centering
        \includegraphics[scale=0.8]{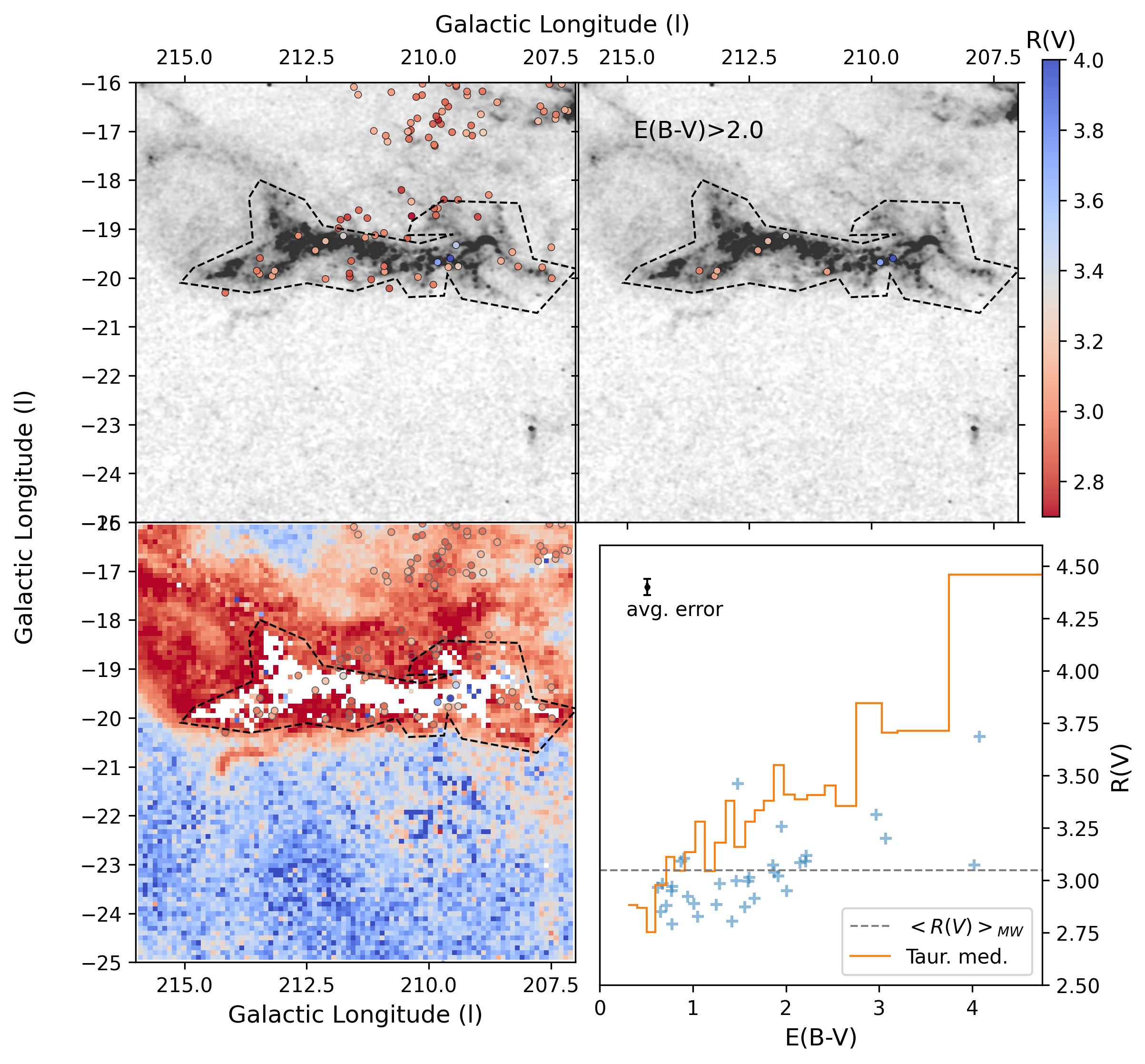}
        \caption{Same as Figure \ref{fig:Taurus4}, but for Orion A.}
    \end{figure*}
    
    \begin{figure*}[h!]
        \centering
        \includegraphics[scale=0.8]{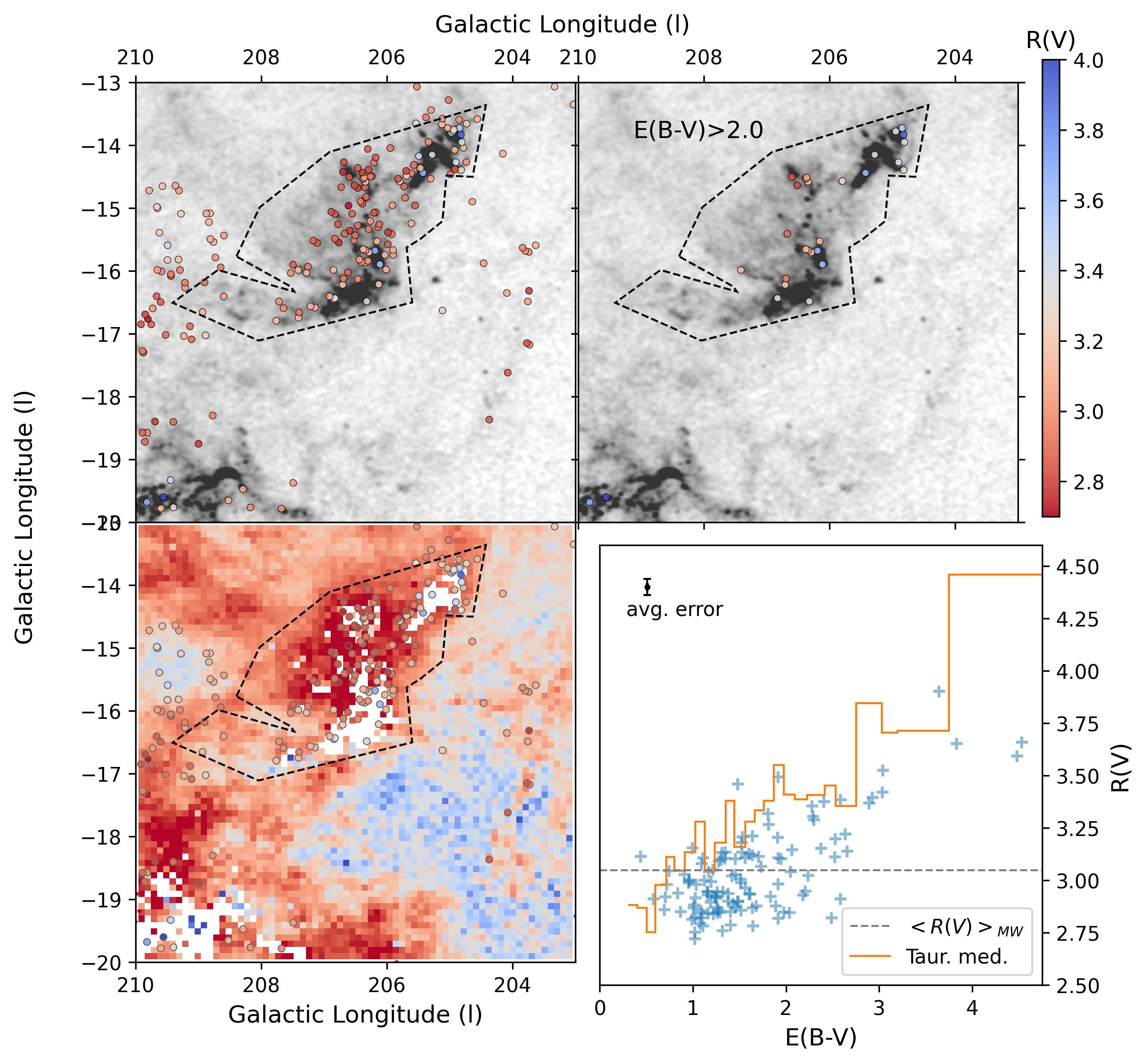}
        \caption{Same as Figure \ref{fig:Taurus4}, but for Orion B.}
    \end{figure*}
    
    \begin{figure*}[h!]
        \centering
        \includegraphics[scale=0.8]{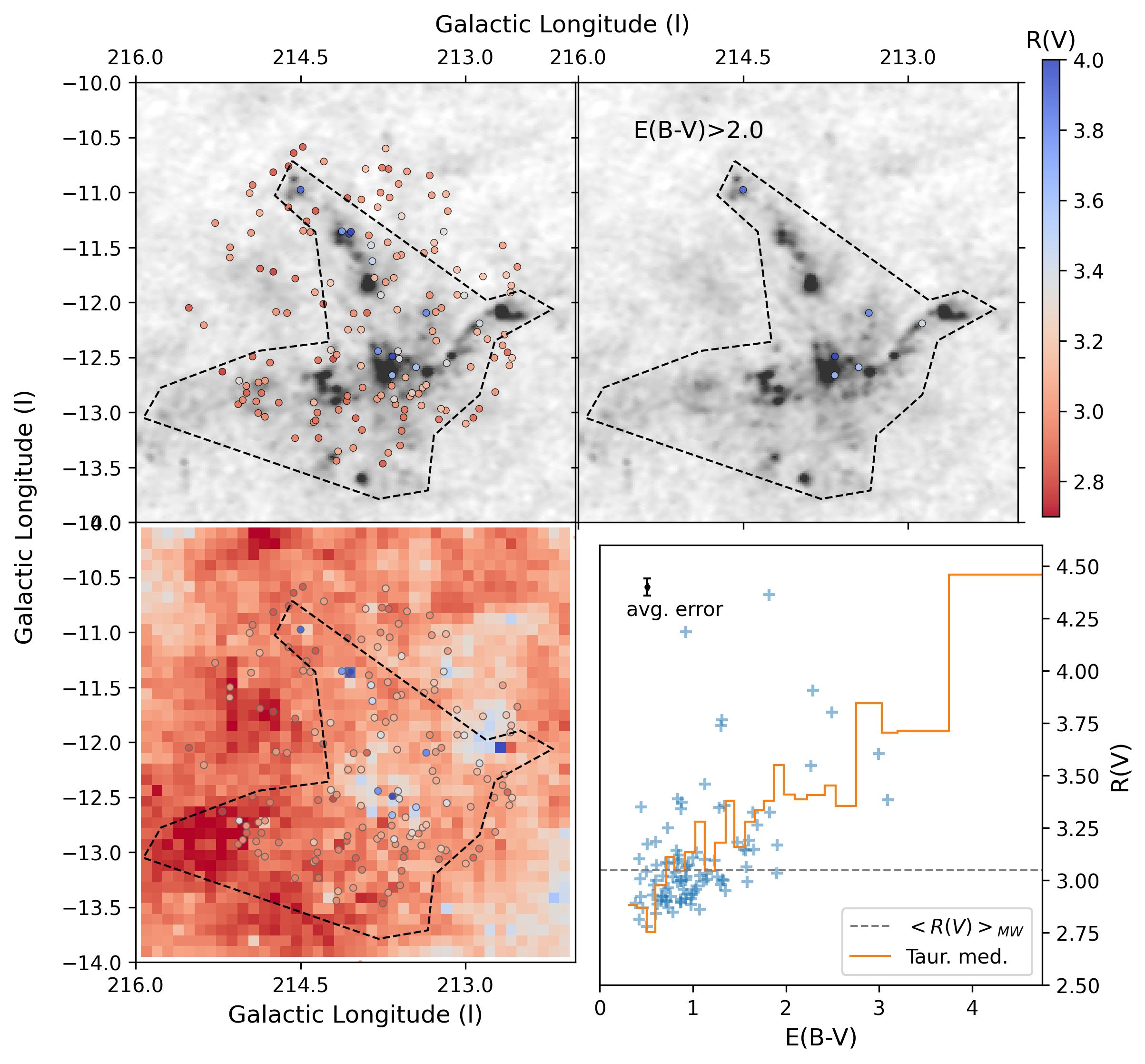}
        \caption{Same as Figure \ref{fig:Taurus4}, but for Monoceros.}
    \end{figure*}

    \begin{figure*}[h!]
        \centering
        \includegraphics[scale=0.8]{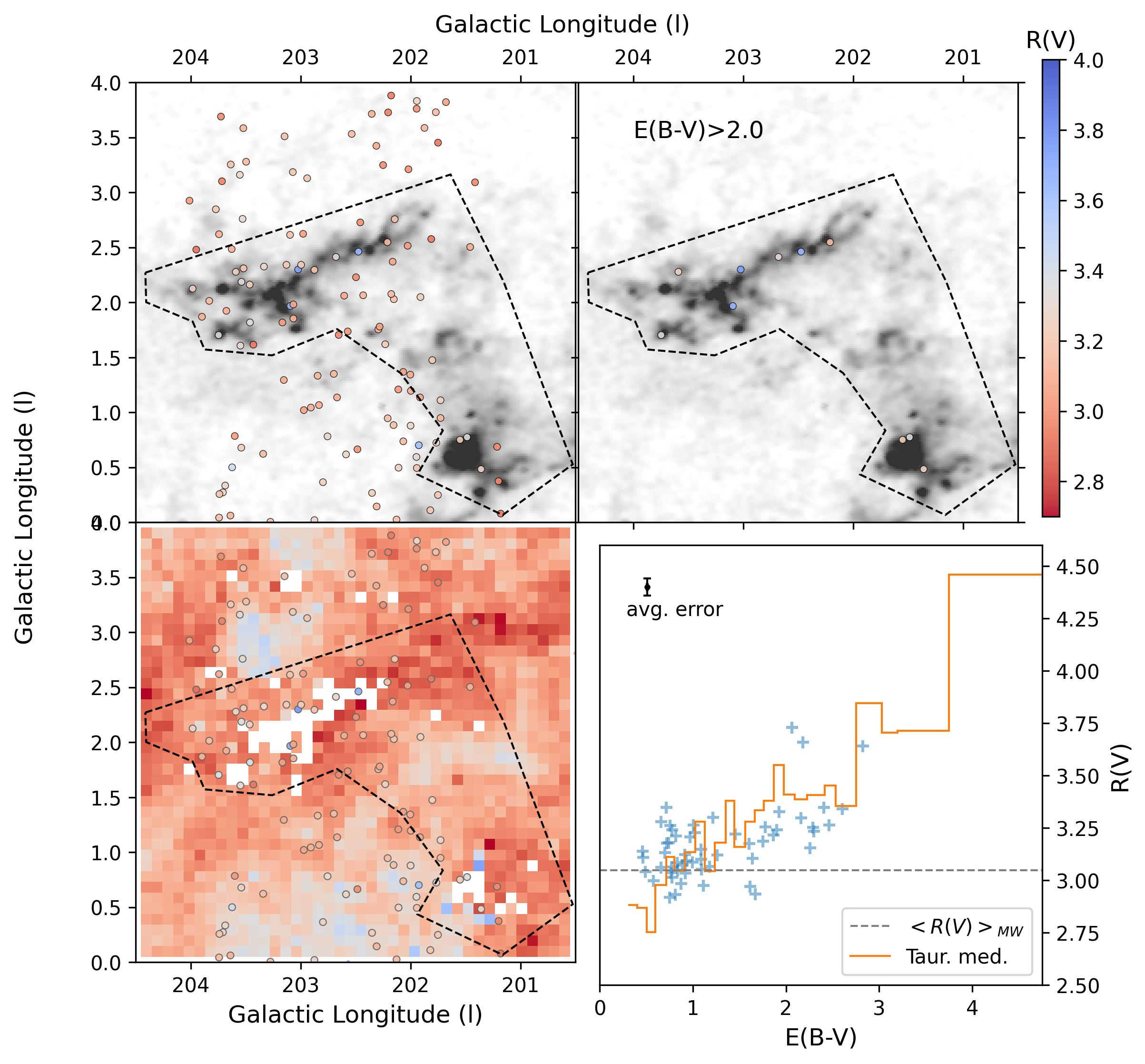}
        \caption{Same as Figure \ref{fig:Taurus4}, but for NGC 2264.}
    \end{figure*}
    
    \begin{figure*}[h!]
        \centering
        \includegraphics[scale=0.8]{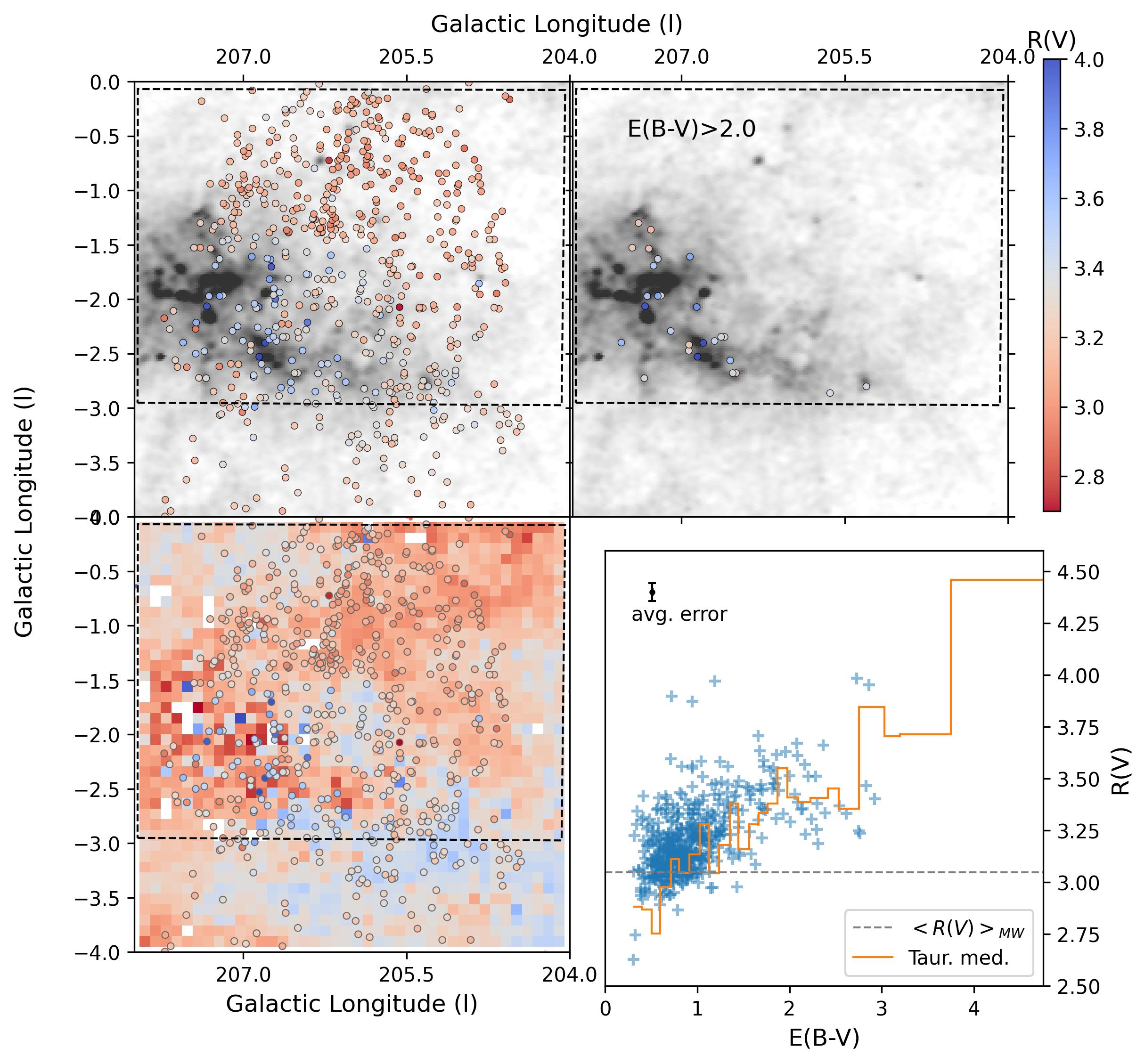}
        \caption{Same as Figure \ref{fig:Taurus4}, but for the Rosette nebula.}
    \end{figure*}

\clearpage
\section{\R\ Residuals for All Clouds} \label{sec:resid_maps}
In the main body of the paper we limited our individual cloud analysis of the to \R\ residuals to clouds with the strongest spatial correlations (Taurus and Ophiuchus). To create these maps, we perform a linear fit through the binned median values, and spatially mapping the residuals ($R(V)_{resid} = \R-\R_{med}$) over the cloud. Here we provide figures of the remaining clouds in our catalog.
    \begin{figure}[h!]
        \centering
        \includegraphics{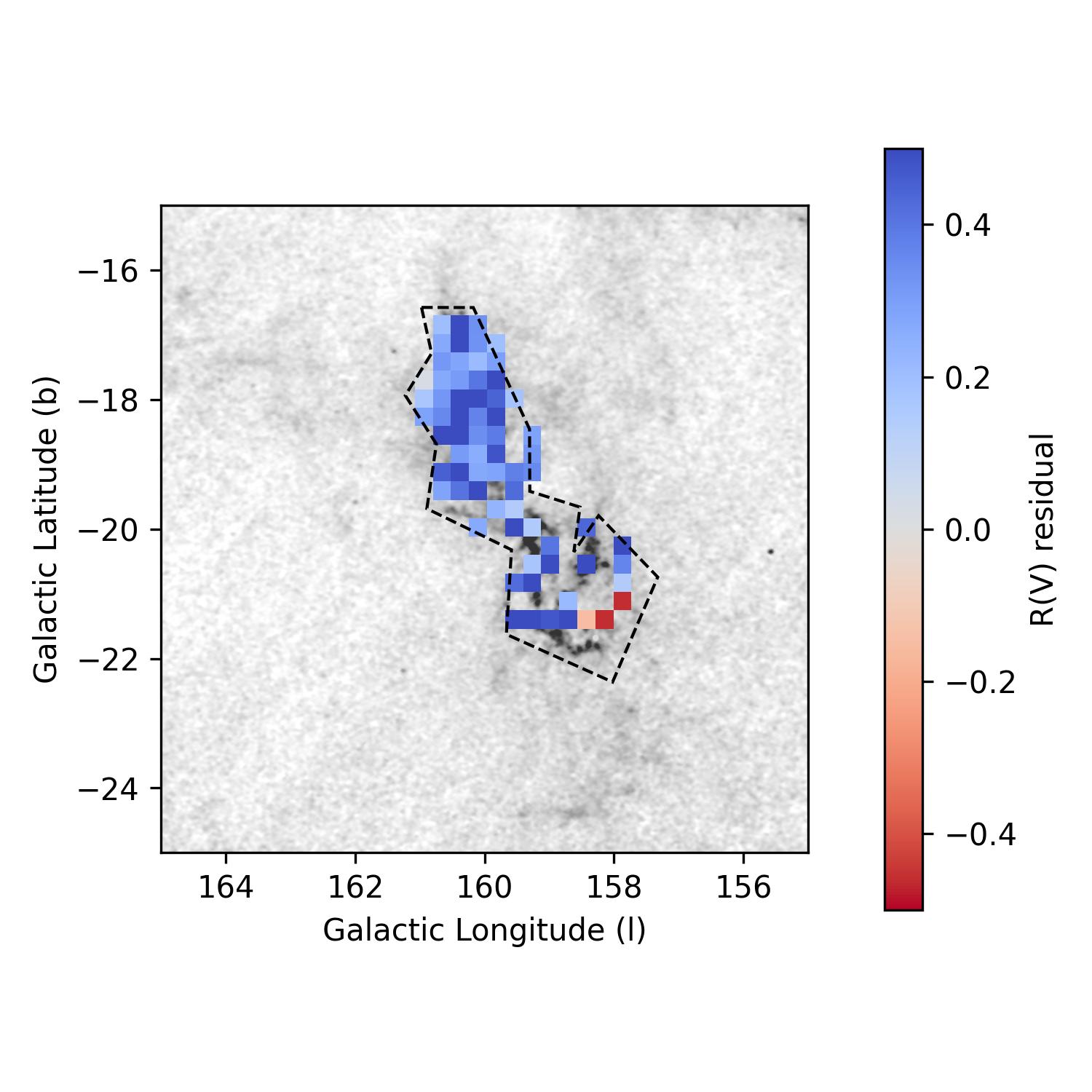}
        \caption{The spatial map of the residual \R\ values for Perseus.}
        \label{fig:resid_map_appendix}
    \end{figure}
    
    \begin{figure}[h!]
        \centering
        \includegraphics{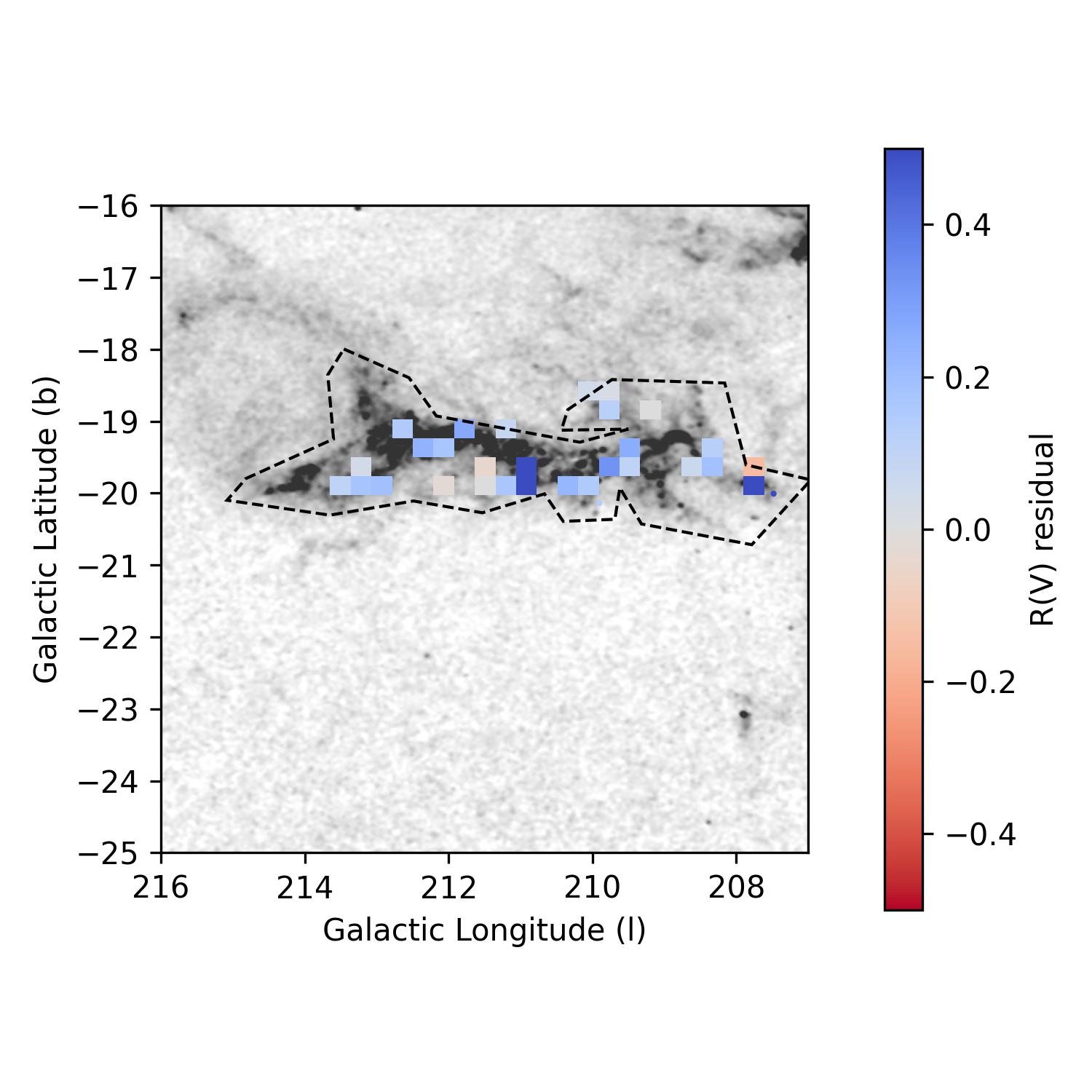}
        \caption{Same as Figure \ref{fig:resid_map_appendix}, but for Orion A.}
    \end{figure}
    
    \begin{figure}[h!]
        \centering
        \includegraphics{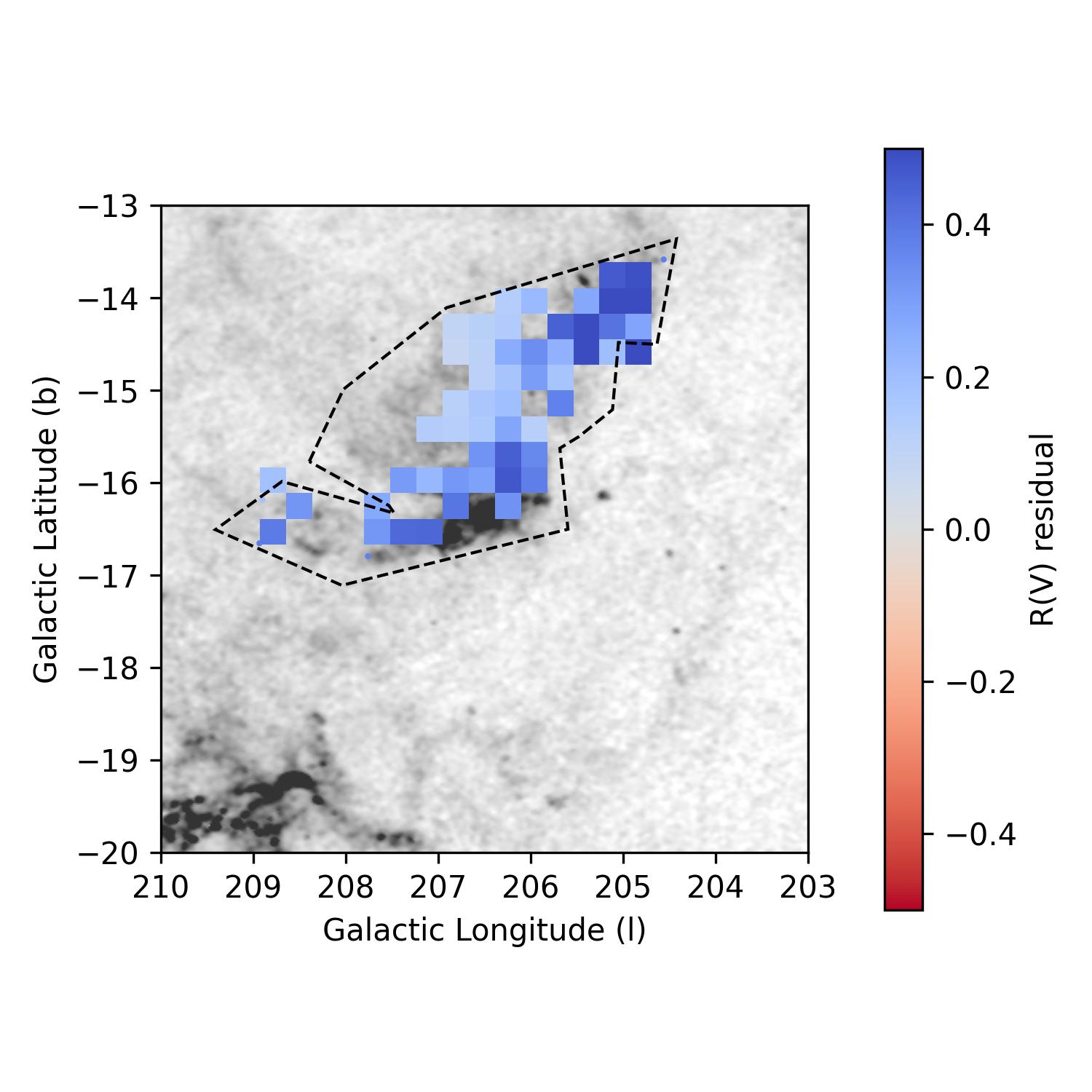}
        \caption{Same as Figure \ref{fig:resid_map_appendix}, but for Orion B.}
    \end{figure}
    
    \begin{figure}
        \centering
        \includegraphics{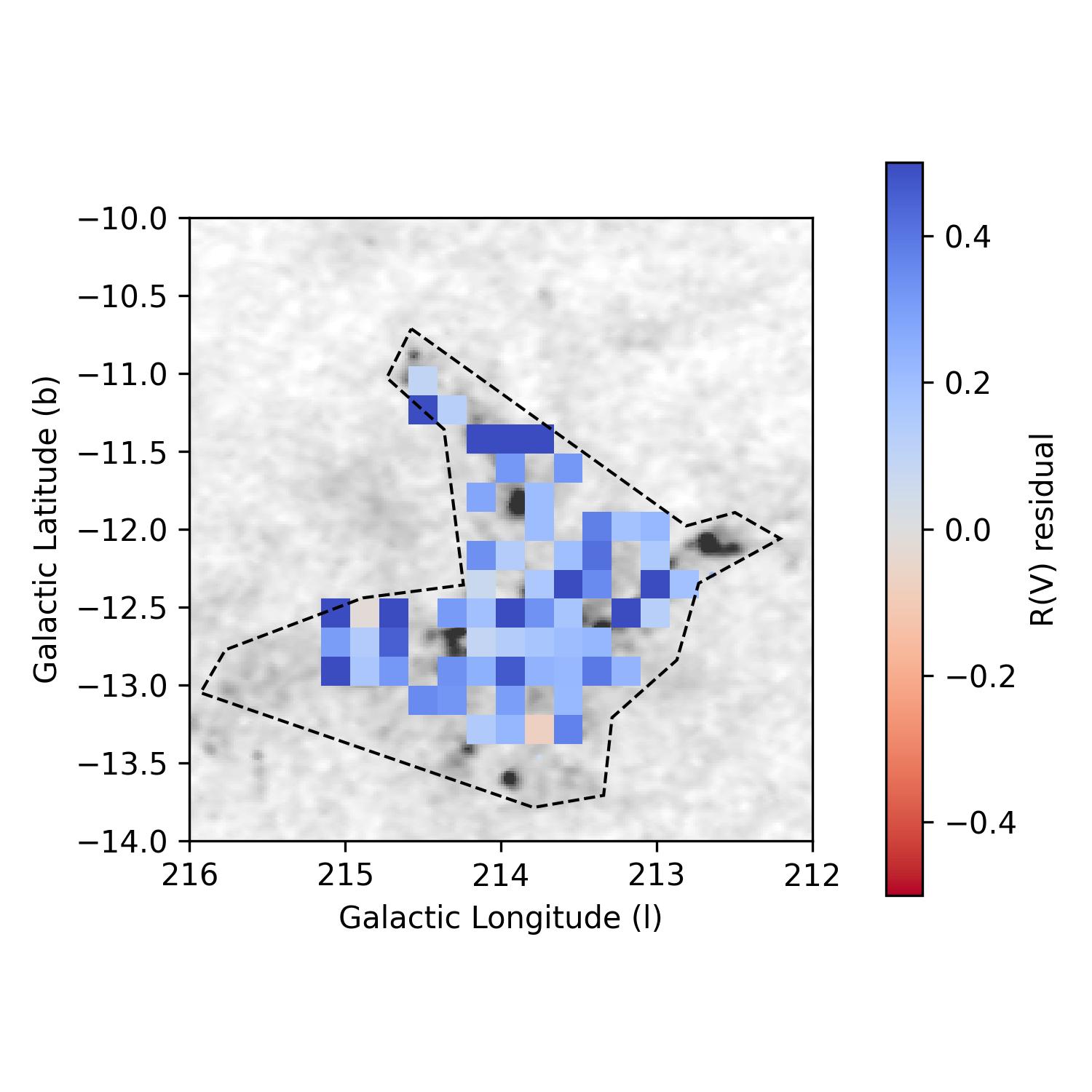}
        \caption{Same as Figure \ref{fig:resid_map_appendix}, but for Monoceros.}
    \end{figure}

    \begin{figure}
        \centering
        \includegraphics{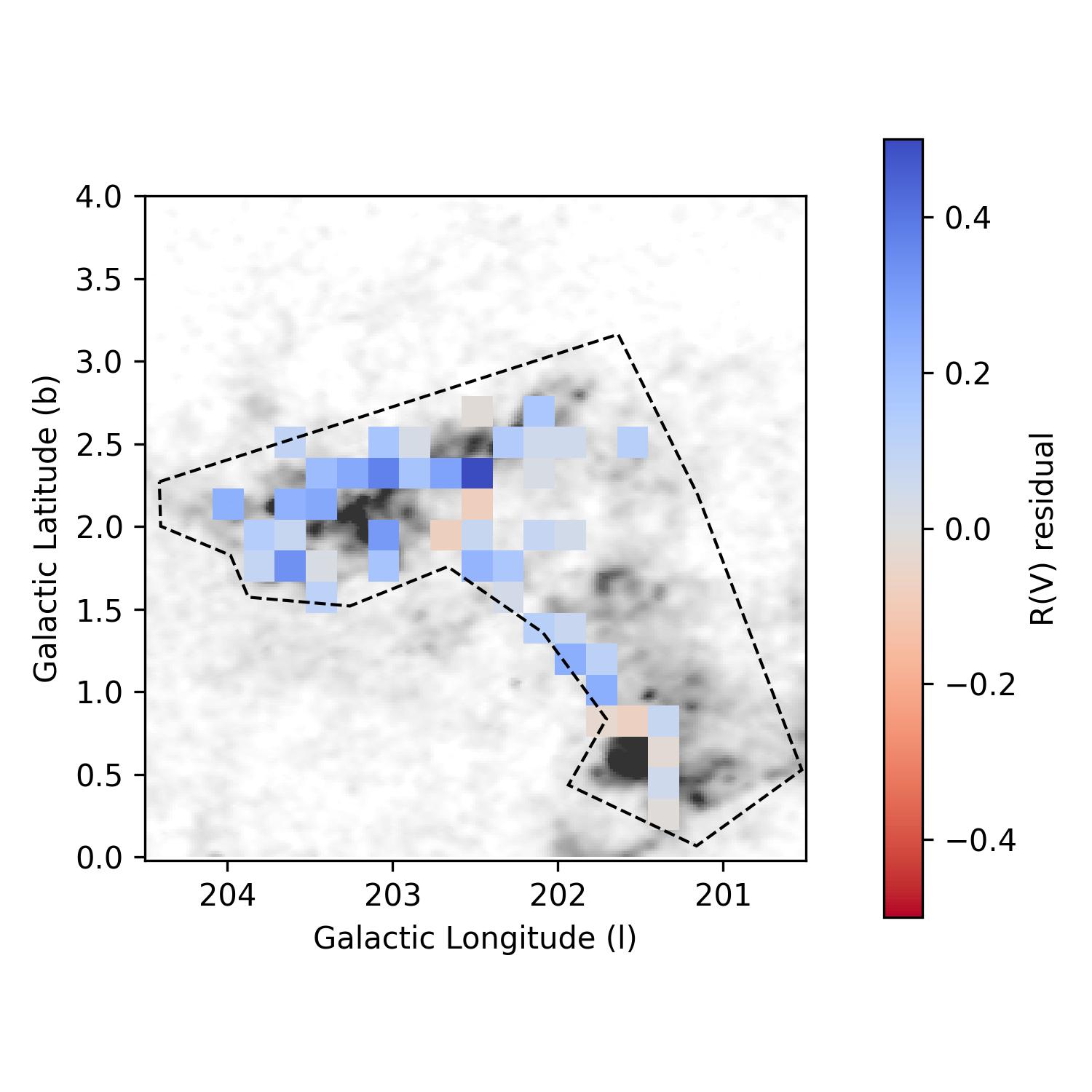}
        \caption{Same as Figure \ref{fig:resid_map_appendix}, but for NGC 2264.}
    \end{figure}
    
    \begin{figure}
        \centering
        \includegraphics{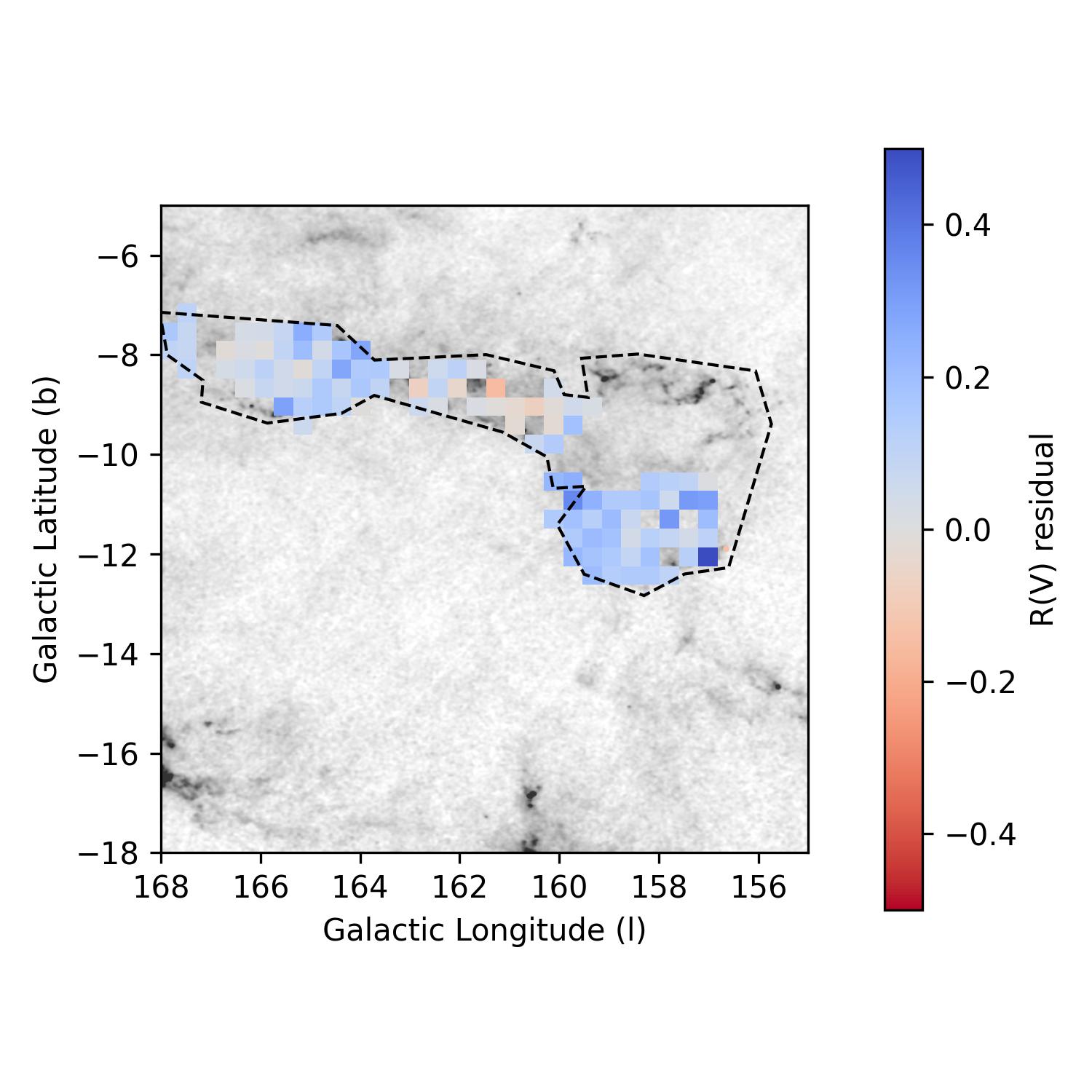}
        \caption{Same as Figure \ref{fig:resid_map_appendix}, but for California.}
    \end{figure}
    
    \begin{figure}
        \centering
        \includegraphics{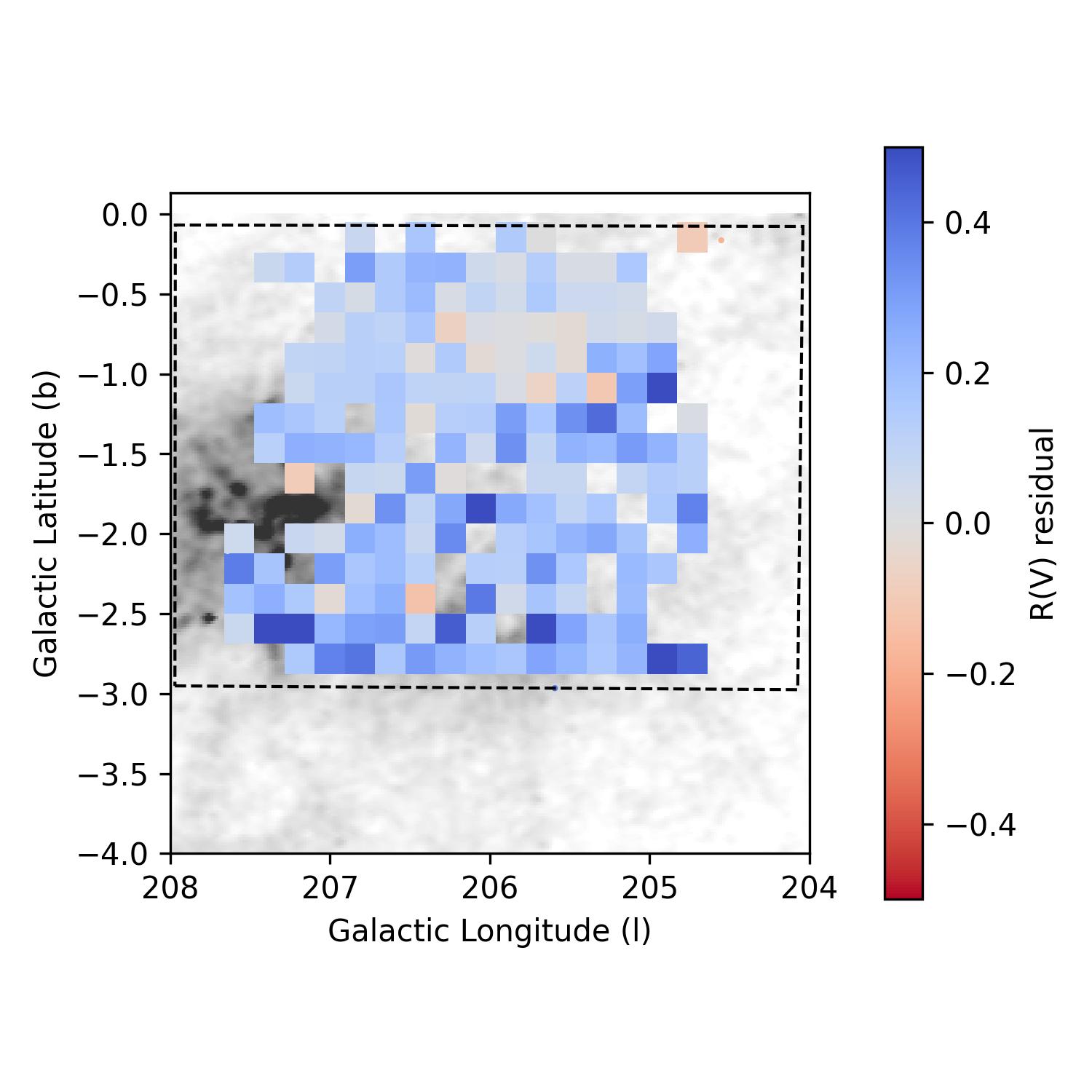}
        \caption{Same as Figure \ref{fig:resid_map_appendix}, but for the Rosette nebula.}
    \end{figure}
\end{document}